\documentclass[review,number,amssymb,amsthm,floatfix]{elsart}
\usepackage{epsfig}
\input epsf
\begin{document}

\begin{frontmatter}
\title{Electromagnetic Shower Properties in a Lead-Scintillator Sampling Calorimeter}
\author[kotwal]{Ashutosh V. Kotwal\thanksref{cor}},  
\author[hays]{Christopher Hays}
\thanks[cor]{Corresponding author. Tel: (919) 660-2563; fax: (919) 660-2525; E-mail address: ashutosh.kotwal@duke.edu}
\address[kotwal]{{\em Physics Department, Duke University, Durham, NC 27708-0305, USA.} }
\address[hays]{{\em Particle Physics Department, Oxford University, Keble Road, Oxford OX1 3RH, UK.} }
\begin{keyword}
calorimeter simulation \sep electromagnetic shower \sep sampling calorimeter \sep energy leakage.
\PACS 07.05.Tp \sep 07.77.Ka  \sep 07.90.+c 
\sep 29.40.-n \sep 29.40.Vj \sep 29.90.+r 
\end{keyword}

\begin{abstract}
The Collider Detector at Fermilab (CDF) is a general-purpose experimental apparatus 
with an inner tracking detector for measuring charged particles, surrounded by a 
calorimeter for measurements of electromagnetic and hadronic showers.  We describe a 
{\sc geant4} simulation and parameterization of the response of the CDF central 
electromagnetic calorimeter (CEM) to incident electrons and photons.  The detector 
model consists of a detailed description of the CEM geometry and material in the 
direction of the incident particle's trajectory, and of the passive material between the 
tracker and the CEM.  We use {\sc geant4} to calculate the distributions of: the 
energy that leaks from the back of the CEM, the energy fraction sampled by the scintillators, 
and the energy dependence of the response.  We parameterize these distributions to 
accurately model electron and photon response and resolution in a custom simulation for 
the measurement of the $W$ boson mass.
\end{abstract}

\end{frontmatter}

\section{Introduction}
\label{sec:intro}
The measurement of the $W$ boson mass with the CDF detector~\cite{wmassPRD} at the Fermilab Tevatron 
 $p \bar{p}$ collider  
 achieves a precision of 
2 parts per 10,000 on the measured energy of electrons from $W$ boson decays~\cite{cdfwmass}.  
A key component of the energy calibration is a detailed simulation of the calorimeter 
response to incident electrons and photons.  This simulation is based on parameterizations 
extracted from {\sc geant4}~\cite{geant} predictions for the distributions relevant to 
the measurement.  In this paper we describe the {\sc geant4} detector model and the 
parameterizations of calorimeter response and resolution.

The electron energy calibration~\cite{cdfwmass} is performed in two steps.  First, the calibrated track 
momentum (with a precision of 1 part per 10,000) is transferred to the measurement of 
calorimeter energy, using the distribution of the ratio of calorimeter energy to the 
track momentum ($E/p$) of electrons from the decays of $W$ and $Z$ bosons.  In the 
second step, the $Z$ boson mass ($m_Z$) is measured using electrons whose cluster 
energy has been calibrated with $E/p$.  After confirming the consistency of the measured $m_Z$ 
with the world average~\cite{pdg}, the $E/p$-based calibration is combined with 
the $m_Z$-based calibration.  

There are several regimes of particle type and energy relevant to this precise calorimeter 
calibration:  the primary electron from the $W$ boson decay, with incident energy in the 
$\approx 20-60$ GeV range; radiated photons from the primary electron, with incident 
energies of $\approx 1$ MeV to $\approx 10$ GeV; and $\approx 0.5-10$ GeV electrons from 
the conversion of photons\footnote{tracks with $p_T < 500$~MeV curl up in the tracker's magnetic field and do not reach the calorimeter.} 
  radiated by the primary electron.  The CEM is a lead-scintillator sampling
 calorimeter. To simulate its 
response to incident electrons and photons of these energies, we parameterize the 
following quantities: 
\begin{itemize}
\item the fraction of the incident particle's energy that leaks out the back of the CEM;
\item the fraction of the deposited energy sampled by the scintillators; 
\item the sampling fluctuations in the scintillator energy fraction; and 
\item the loss of response due to absorption and back-scatter of low-energy particles.
\end{itemize}

In the following we describe the detailed detector geometry implemented in {\sc geant4} 
(Section~\ref{sec:geantGeometry}); the fractional energy leakage for a given particle 
type and energy (Section~\ref{sec:leakage}); the sampling fraction and resolution of the 
calorimeter (Section~\ref{sec:samplingmodel}); and the absorption of energy in the 
passive material that results in a non-linear calorimeter response 
(Section~\ref{sec:nonlinearity}).

\section{Detector Model}
\label{sec:geantGeometry}
\begin{figure*}[!htp]
\begin{center}
\includegraphics[width=5.7in]{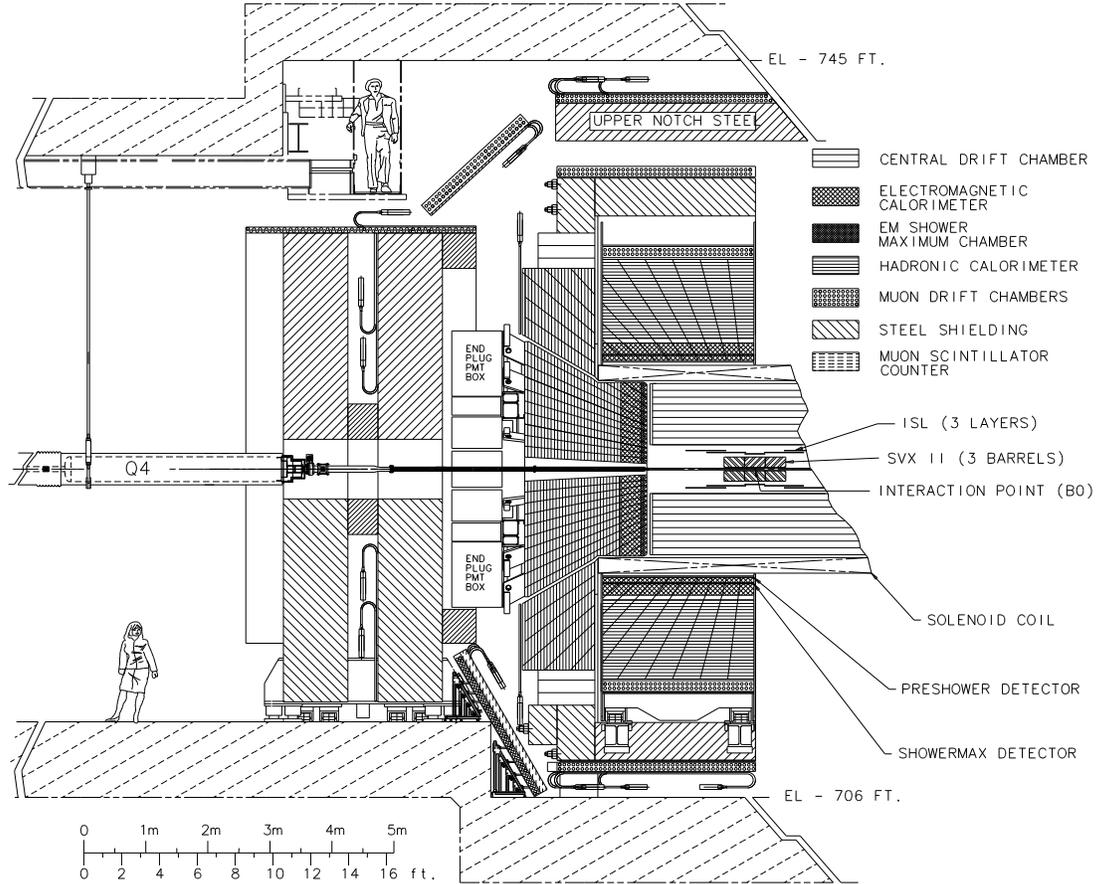}
\caption{A cut-away view of the CDF detector.  We use a simplified {\sc geant4} model 
of the outer casing of the central drift chamber, the time-of-flight detector (not shown), 
the solenoid coil, the preshower detector, and the central electromagnetic calorimeter. }
\label{fig:cdf}
\end{center}
\end{figure*}

The CDF detector~\cite{wmassPRD,cdfNIM,jpsiReference,xsecPRD}  is shown in Fig.~\ref{fig:cdf}.  The detector model implemented in the 
{\sc geant4} simulation includes the components from the outer radius of the central tracking drift 
chamber~\cite{cot} to the back end of the CEM calorimeter.  These components are the outer aluminum 
casing of the tracker, the time-of-flight (TOF) system~\cite{tof} attached to this casing, the solenoidal 
coil~\cite{solenoid} that provides a nearly uniform 1.4~T magnetic field in the tracking volume, the 
central preshower system (CPR)~\cite{cpr} beyond the solenoid, and the CEM calorimeter (including 
longitudinal segmentation)~\cite{cemnim}.

\begin{figure*}[!tbhp]
\begin{center}
\includegraphics[width=4.in]{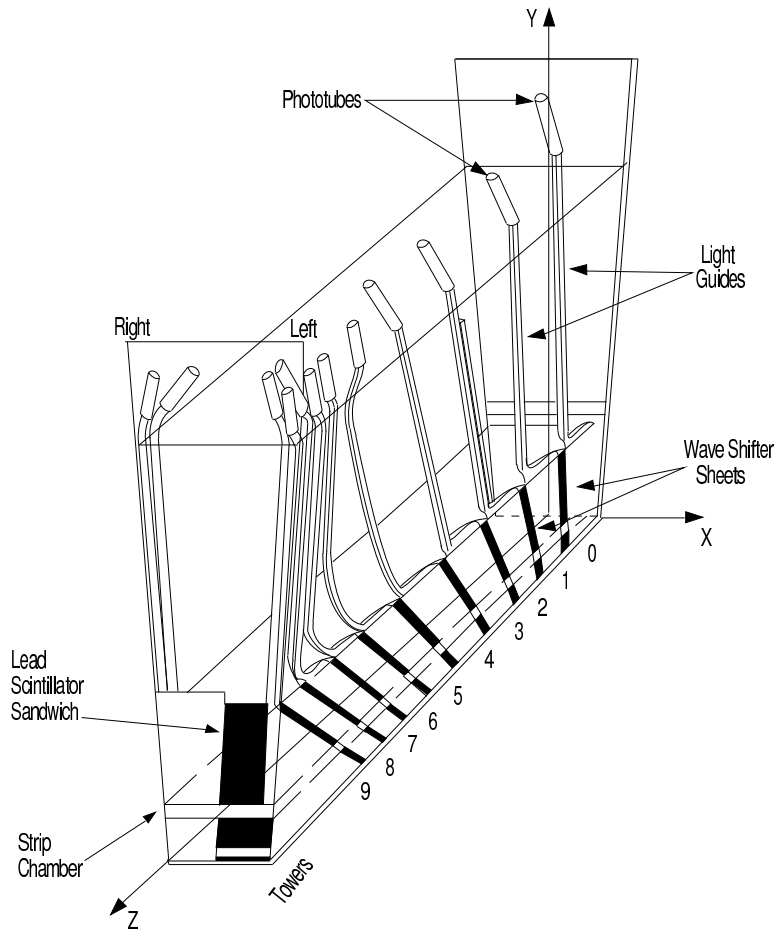}
\caption{The CEM calorimeter, with tower segmentation shown. The $z$-axis points along the beam direction towards
 higher pseudorapidity ($\eta$), and the $x$-axis points in the azimuthal direction.  }
\label{fig:cem}
\end{center}
\end{figure*}

The CEM calorimeter is divided into $0.1 \times 0.15~\eta-\phi$~\cite{conventions} towers, 
shown in Fig.~\ref{fig:cem}.  
 The tower geometry depends on $\eta$, with 
towers numbered according to their distance in $\eta$ from $\eta=0$.  The longitudinal 
segmentation of Tower 0 is an alternating system of 31 scintillator sheets and 30 
aluminum-clad lead sheets, with a plate of aluminum at the front end of the tower.  Each 
lead sheet is 3.175 mm thick and the aluminum cladding is 380 $\mu$m thick on each side 
of the sheet.  Each scintillator sheet is 5 mm thick.  A thin (6 mm) aluminum casing 
contains a strip and wire chamber at the position of shower maximum (after six 
lead-scintillator sandwiches).  Almost all the material between the tracking volume and the first 
scintillator -- the outer casing of the tracker, the solenoid coil and the CEM front 
plate -- is aluminum.  The TOF and CPR are a combination of scintillator and aluminum. 
Table~\ref{tbl:cem} shows the materials in the {\sc geant4} model with their 
thicknesses in units of mm ($x$) and radiation length ($x_0 \equiv x/X_0$).  

\begin{table*}
\begin{center}
\begin{tabular}{lcc}
\hline
\hline
Material & Thickness $x$ (mm)	& $x_0$ \\
\hline
CEM lead sheet	        & $3.175 \times 30$ & $0.5658 \times 30 = 16.974$ \\
CEM aluminum cladding   & $0.76 \times 30$  & $0.0085 \times 30 = 0.255$ \\
CEM scintillator sheet	& $5.0 \times 31$   & $0.0121 \times 31 = 0.375$ \\
CES aluminum	        & 6.0	& 0.07 \\
Solenoid coil aluminum	& 76.5	& 0.86 \\
CEM front-plate aluminum& 14.0	& 0.157	\\
Tracker+TOF+CPR aluminum& 27.0	& 0.303 \\
\hline\hline
\end{tabular}
\caption{The materials and their thicknesses in the {\sc geant4} implementation 
of the CEM and upstream materials~\cite{cemnim}.  }
\label{tbl:cem}
\end{center}
\end{table*}

We model the material between the tracking volume and the first CEM scintillator as an aluminum 
plate of 6.51 cm thickness plus an aluminum-clad lead sheet at the front of the 
active calorimeter volume.  The additional lead sheet is included for simplicity: 
with this sheet the CEM volume is modelled as 31 alternating lead-scintillator layers.  
Combined, the 6.51 cm of aluminum and the single lead sheet reproduce the total 
radiation lengths upstream of the first scintillator layer.  

The geometry of other towers is implemented according to Table~\ref{tbl:cemtowers}.  
As $|\eta|$ increases, the number of lead sheets in a tower decreases, compensating 
for the increasing path length.  This approximately maintains the same total number 
of radiation lengths 
traversed by a particle originating from the center of 
the detector.  For each removed lead sheet, acryllic is used in its place and the 
subsequent scintillator sheet is blackened so that the sampling fraction is unaffected.  
In the {\sc geant4} model described here we neglect the additional radiation lengths 
contributed by acryllic and blackened scintillator.  In the CDF measurement of the 
$W$ boson mass~\cite{cdfwmass} a correction is applied to account for this extra plastic material.  

\begin{table*}[!th]
\begin{center}
\begin{tabular}{ccc}
\hline
\hline
Tower & Thickness ($x_0$) & Number of lead sheets \\
\hline
0       & 17.9	& 30 \\
1       & 18.2	& 30 \\
2       & 18.2  & 29 \\
3       & 17.8	& 27 \\
4       & 18.0	& 26 \\
5       & 17.7	& 24 \\
6    	& 18.1	& 23 \\
7      	& 17.7	& 21 \\
8      	& 18.0	& 20 \\
\hline\hline
\end{tabular}
\caption{The thicknesses of the CEM towers~\cite{cemnim}, in units of radiation length.  The 
aluminum front-plate is included in this calculation, but the other upstream material is not. 
}
\label{tbl:cemtowers}
\end{center}
\end{table*}

\section{Longitudinal Leakage}
\label{sec:leakage}
The entire assembly of the lead-scintillator sandwich plus upstream material is about 18 
radiation lengths thick.  A typical 50 GeV incident electron will deposit about 48 GeV of 
its energy in this structure, and  about 5\% will leak out the back.  The leakage 
energy fluctuates from shower to shower, contributing to the measurement resolution of the 
electron's energy.  Because this resolution is not gaussian, it is important to model the 
leakage distribution directly.  We develop parameterizations of the energy loss distributions 
of incident electrons and photons as functions of incident energy and angle, and of 
calorimeter thickness.

\subsection{Electron Leakage Model}
\label{sec:electronLeakage}
\begin{figure}
\begin{center}
\includegraphics[width=3.3in]{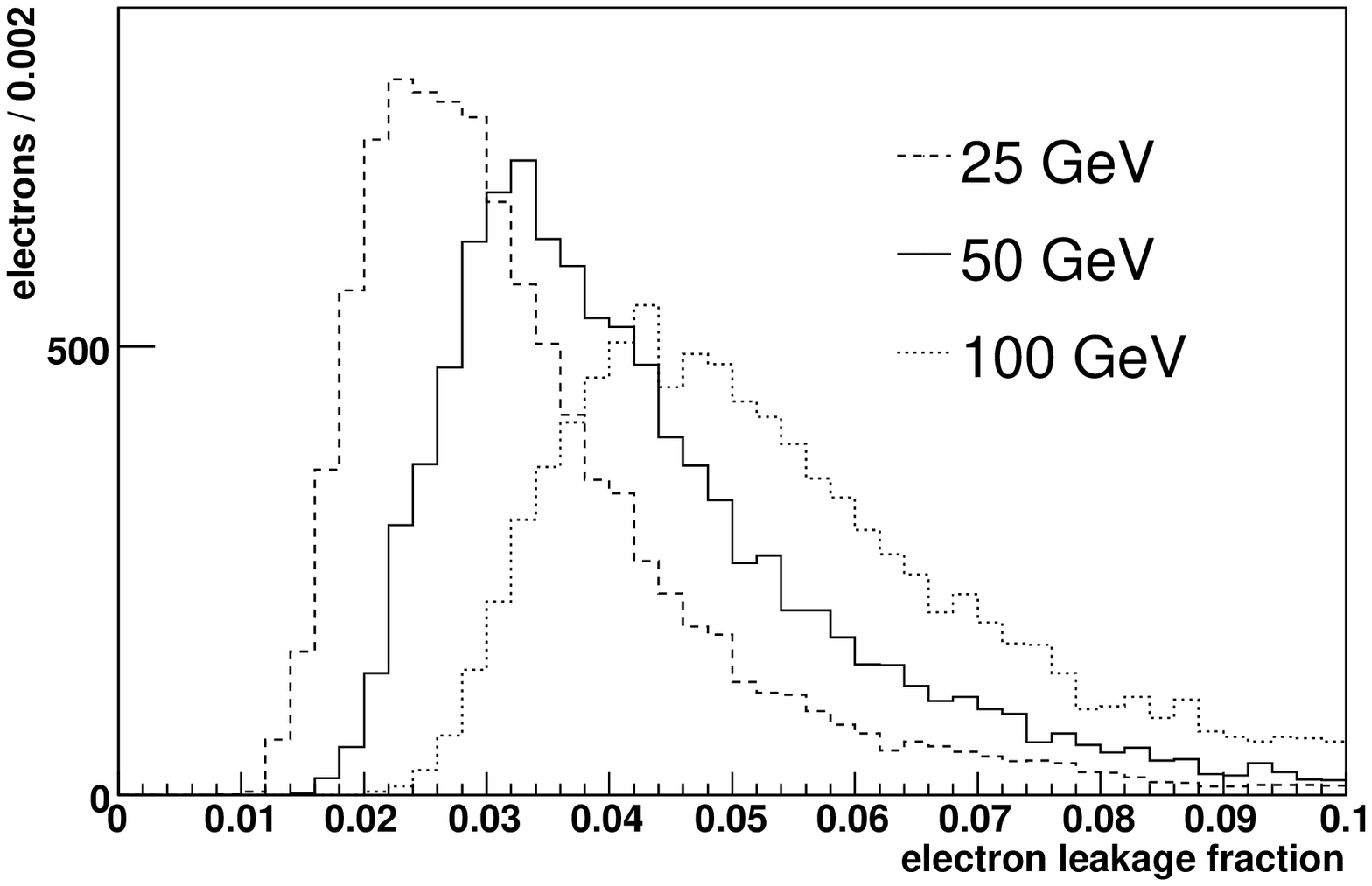}
\includegraphics[width=3.3in]{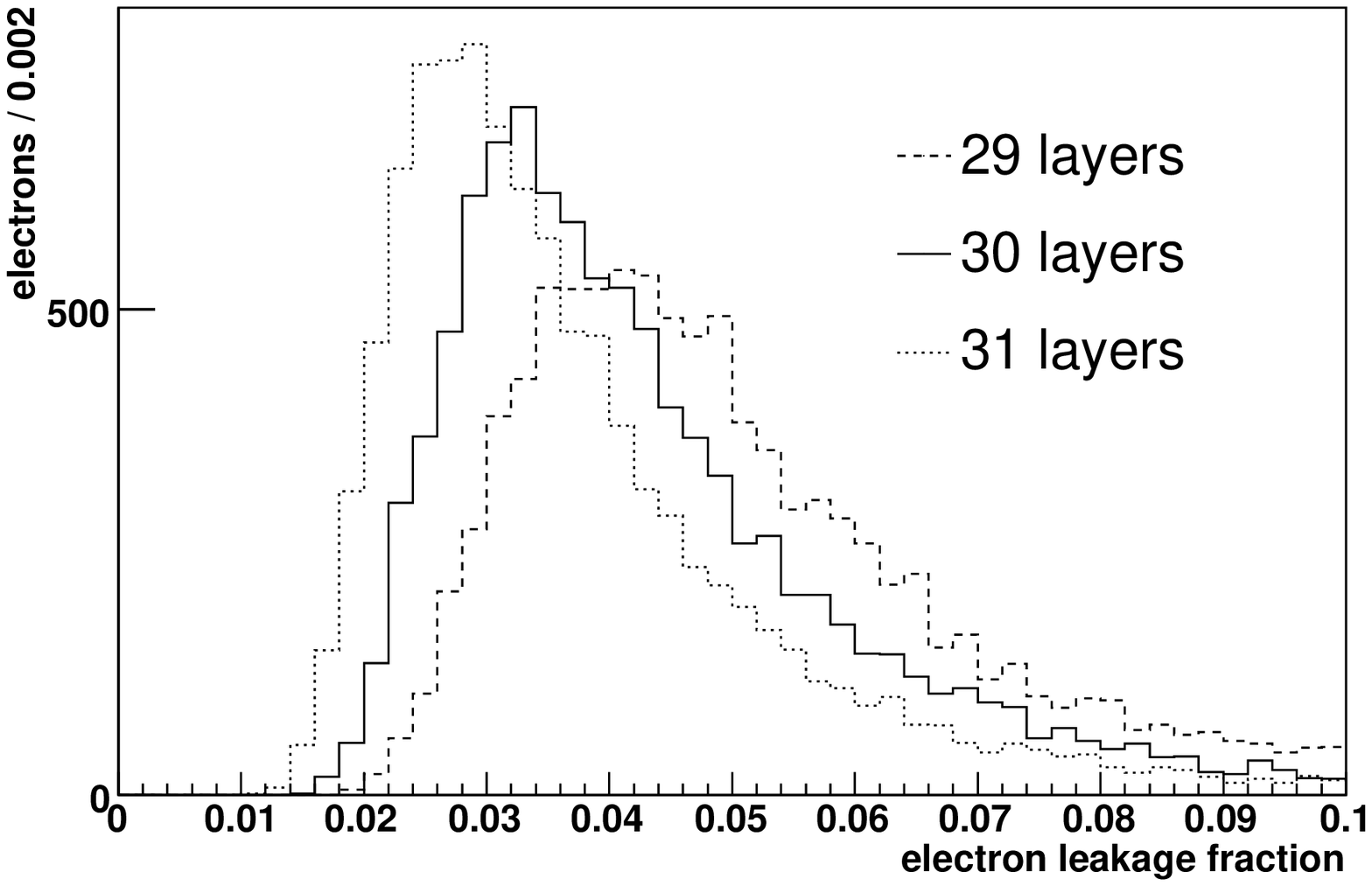}
\end{center}
\caption{Longitudinal energy leakage fraction $f_l$ for electrons of different beam energies (top), and for CEM thicknesses varying by $\pm 1$ lead/scintillator layer (bottom). These distributions have been calculated using our
 {\sc geant4} model.  }
\label{linearLeakageVsEnergyThicknessPlots}
\end{figure}

\begin{figure}
\begin{center}
\includegraphics[width=3.2in]{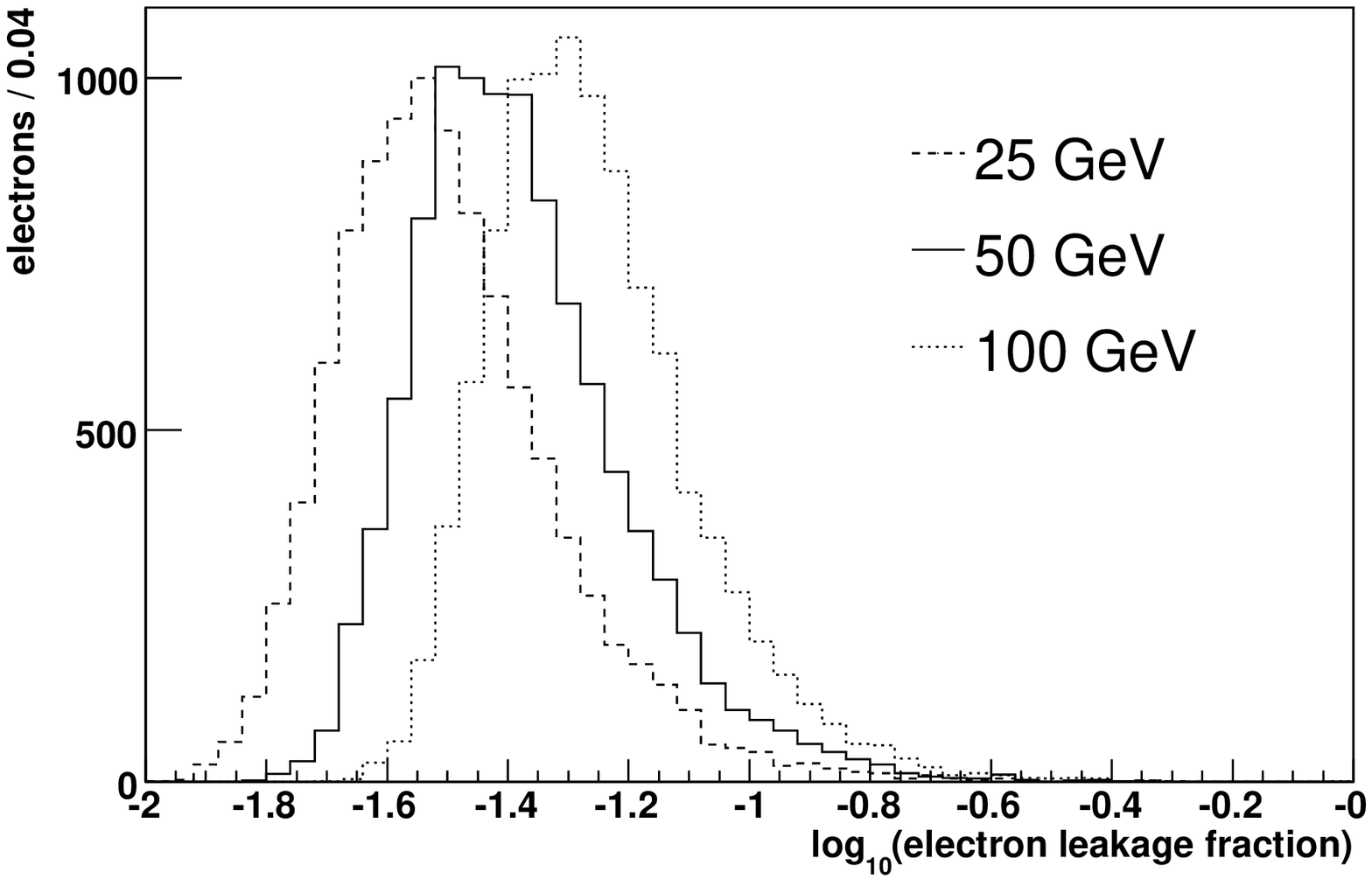}
\includegraphics[width=3.2in]{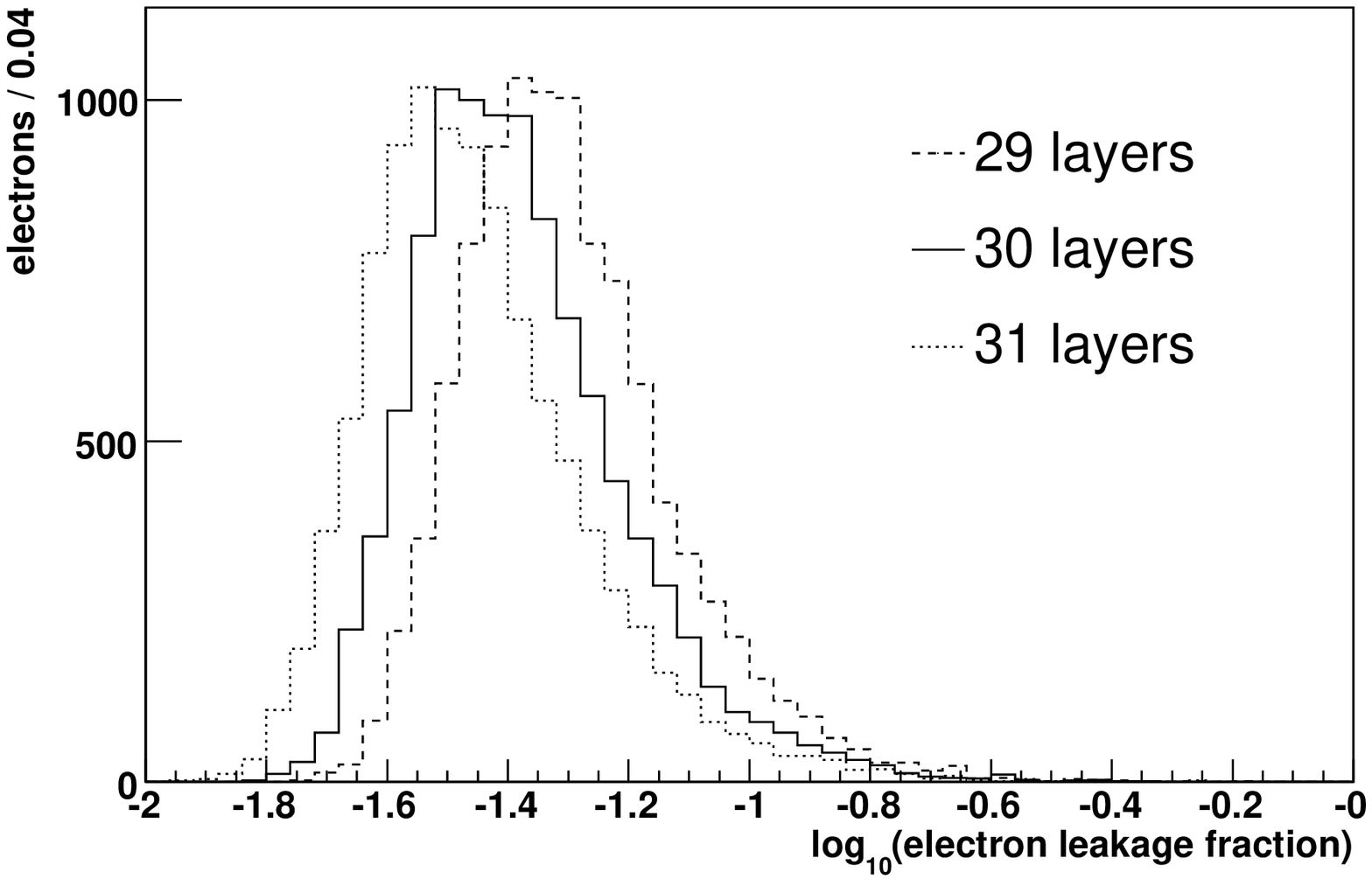}
\end{center}
\caption{The logarithm of the longitudinal energy leakage fraction $f_l$ for incident electrons 
with energies of 25, 50 and 100 GeV (top), and for CEM thicknesses decreased and increased 
by one lead-scintillator layer (bottom) with respect to the nominal value of 30 lead sheets 
in Tower 0. These distributions have been calculated using our
 {\sc geant4} model. }
\label{fig:logLeakagevsEandx0}
\end{figure}

We define the leakage fraction $f_l$ as the ratio of longitudinal leakage energy to the 
incident energy of the electron, where the incident energy is defined as the energy of the 
electron as it enters the outer casing of the drift chamber.  Figure~\ref{linearLeakageVsEnergyThicknessPlots} shows the 
leakage fraction for three incident energies (25, 50, and 100 GeV) and for three CEM thicknesses 
(with 29, 30, and 31 lead-scintillator layers) for 50 GeV  electrons.  The 
distributions are skewed due to a tail that extends beyond 10\%.  Because of this skew it is more 
convenient to study the logarithm of the leakage fraction.  Figure~\ref{fig:logLeakagevsEandx0} 
shows this distribution for the same incident energies and thicknesses as 
Fig.~\ref{linearLeakageVsEnergyThicknessPlots}.  We find that $\log_{10} f_l$ has a nearly invariant shape, 
with the peak position shifting as a function of incident energy and CEM thickness.  We use 
this feature to devise a compact parameterization of the leakage fraction. 

\begin{figure}
\begin{center}
\begin{minipage}{0.475\textwidth}
\includegraphics[width=2.7in]{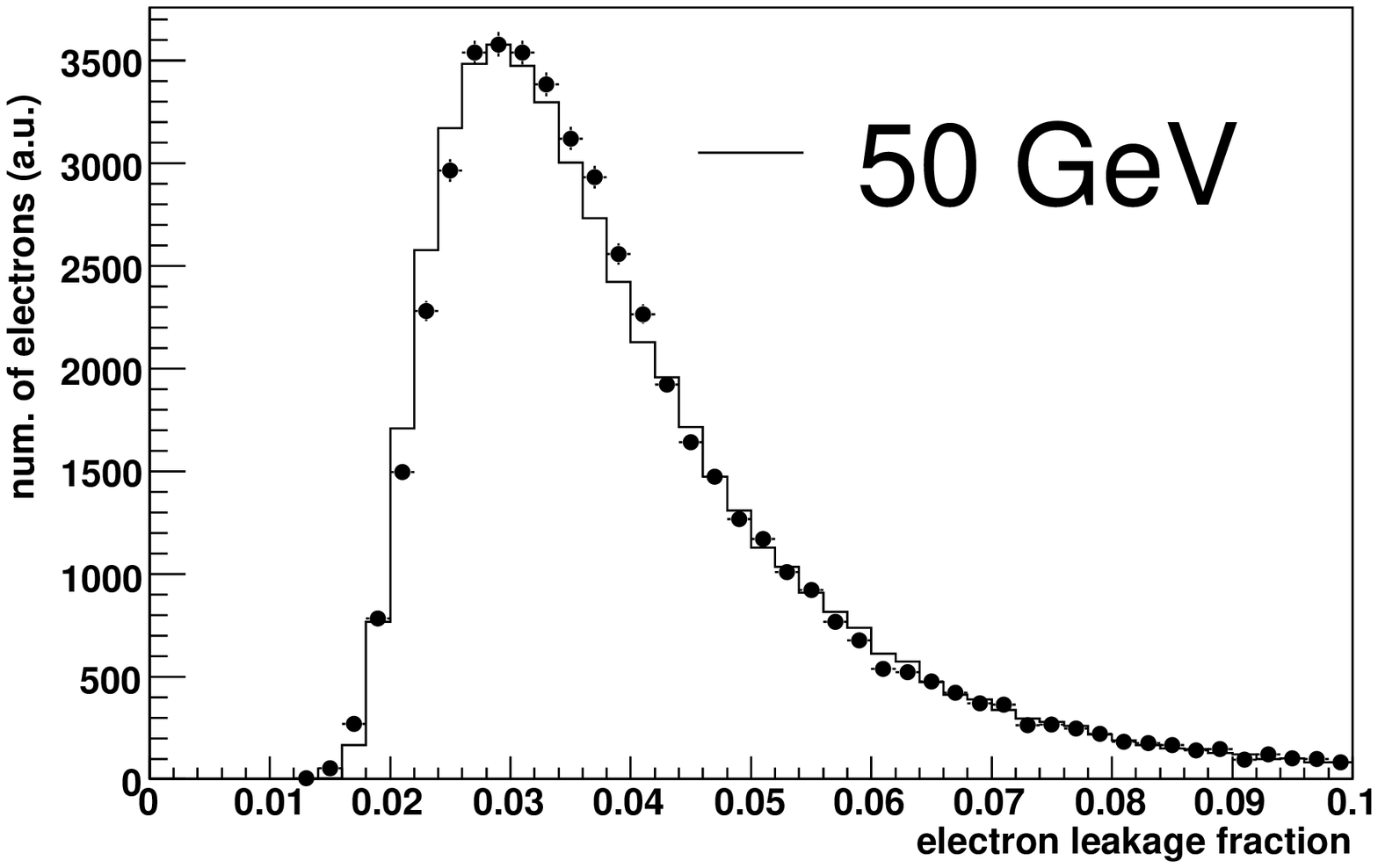}
\end{minipage}
\end{center}
\caption{A comparison of the {\sc geant4} model calculation (points) and our parameterization (histogram) 
of the distribution of the energy leakage fraction $f_l$ for 50 GeV electrons. }
\label{fig:compareGeant1}
\end{figure}

As shown in Fig.~\ref{fig:compareGeant1} for electrons with 50 GeV of energy normally incident on Tower 0, the standard Gamma 
distribution $\Gamma(x) = x^a e^{-x}$ gives a good description of the distribution of $\log_{10} f_l$, with 
$x \equiv (\log_{10} f_l - \mu) / \sigma$.  At this energy, the parameters of the Gamma 
distribution are $a=5$, $\sigma=0.07$ and $\mu= -1.8$.  Of these parameters, only 
$\mu = \left< \log_{10} f_l \right>$ depends on the incident electron energy $E$ and CEM 
thickness $x_0$ at sufficiently large energy.  The following parameterization accurately 
models this dependence for $E > 10$ GeV:
\begin{eqnarray}
 \Delta \mu = (\Delta \log_2 E - \Delta x_0) / 8, 
\label{leakageModelEqn}
\end{eqnarray}
implying that a change in the incident energy by a factor of two has the same effect on the leakage energy distribution as a change in the calorimeter thickness by one radiation length. 

\begin{figure*}
\begin{center}
\begin{minipage}{0.475\textwidth}
\includegraphics[width=2.7in]{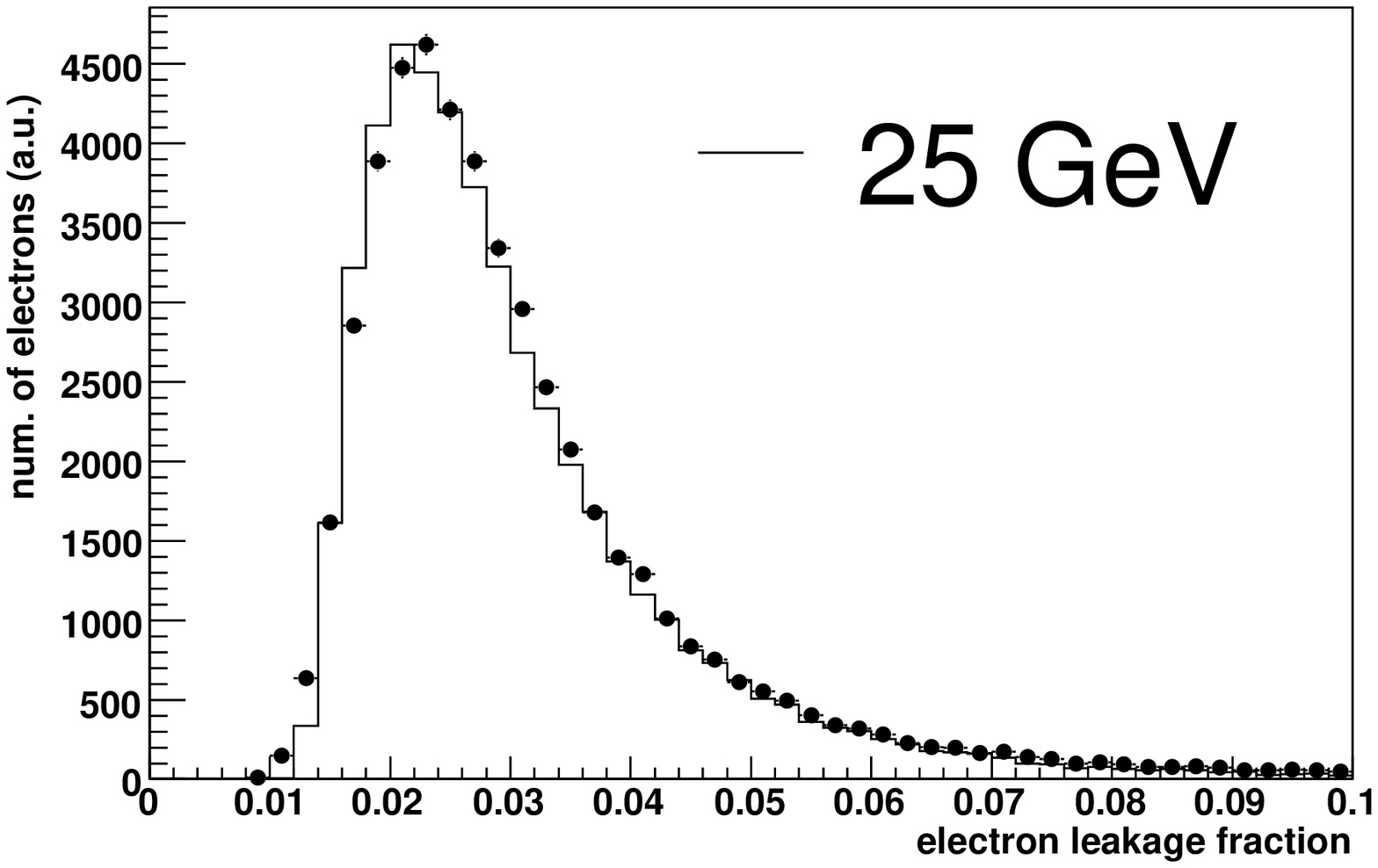}
\end{minipage}
\begin{minipage}{0.475\textwidth}
\includegraphics[width=2.7in]{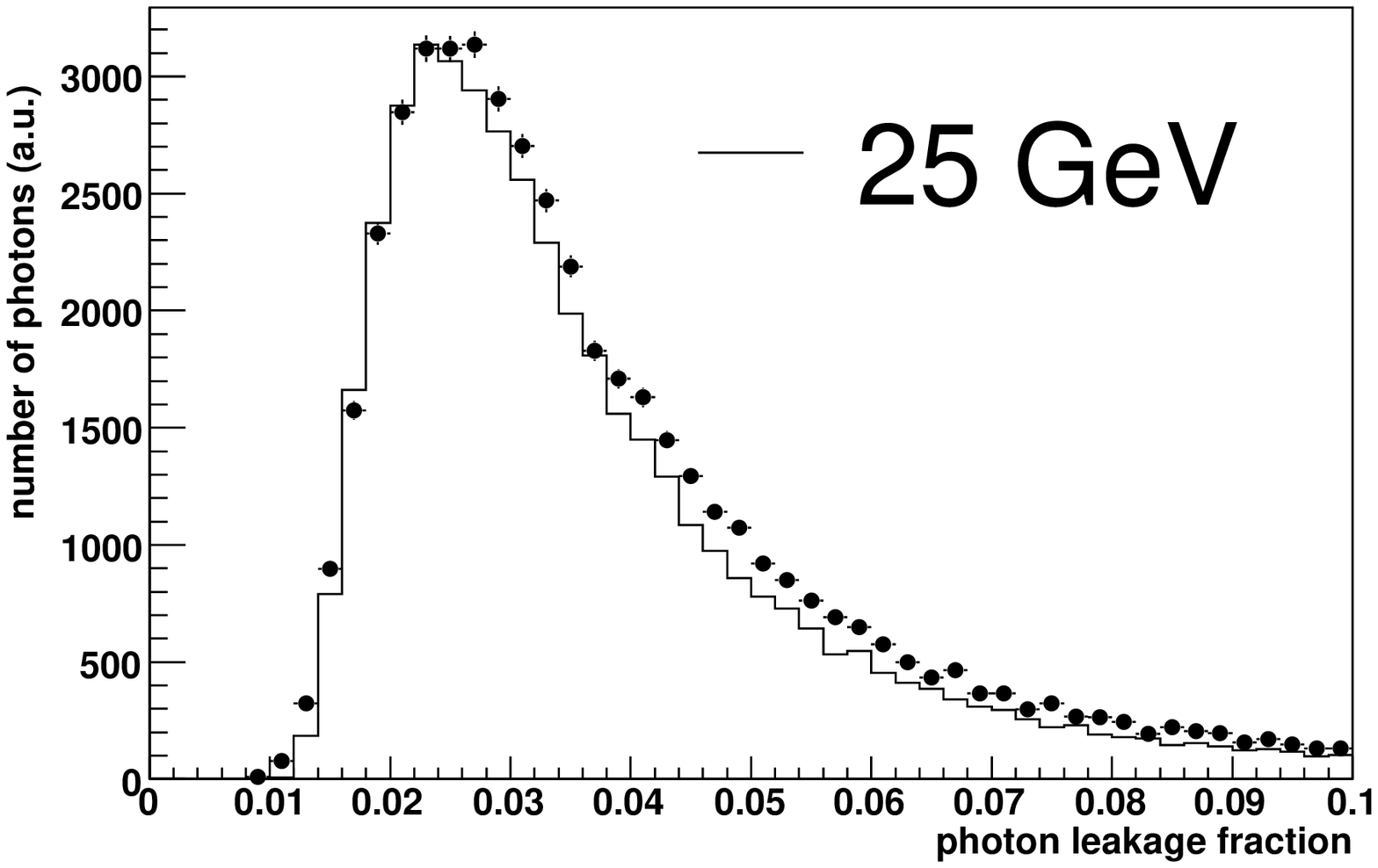}
\end{minipage}
\begin{minipage}{0.475\textwidth}
\includegraphics[width=2.7in]{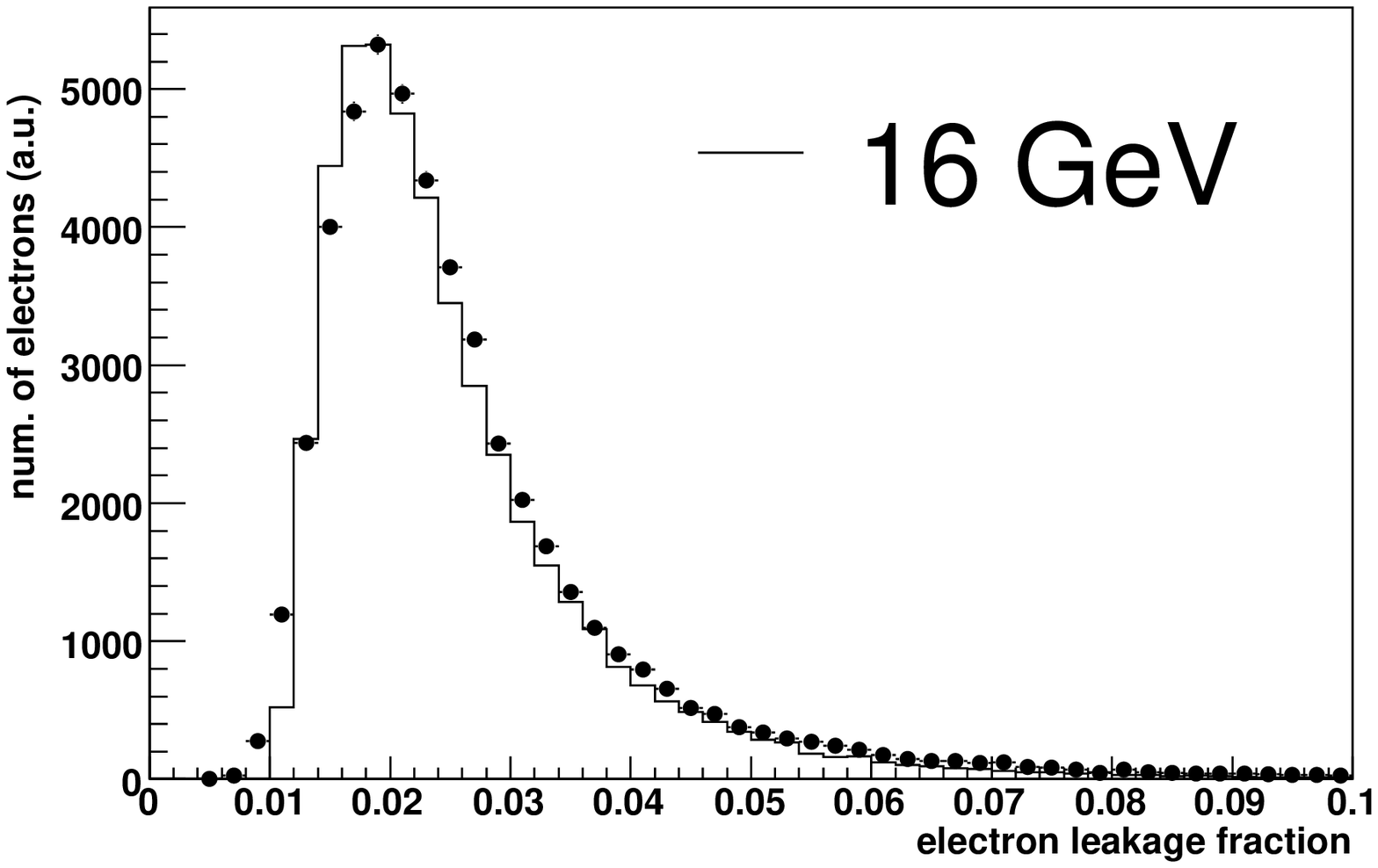}
\end{minipage}
\begin{minipage}{0.475\textwidth}
\includegraphics[width=2.7in]{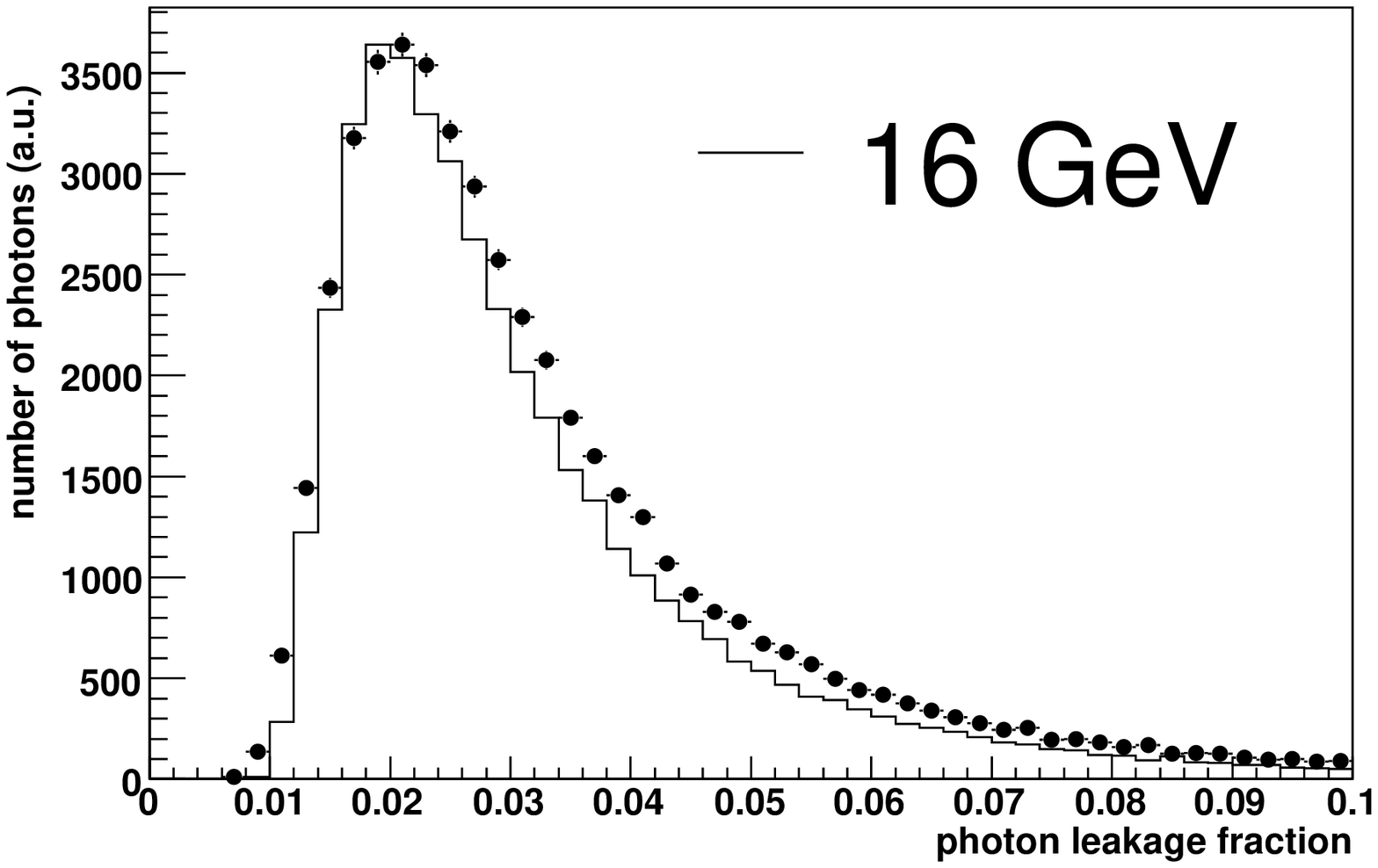}
\end{minipage}
\begin{minipage}{0.475\textwidth}
\includegraphics[width=2.7in]{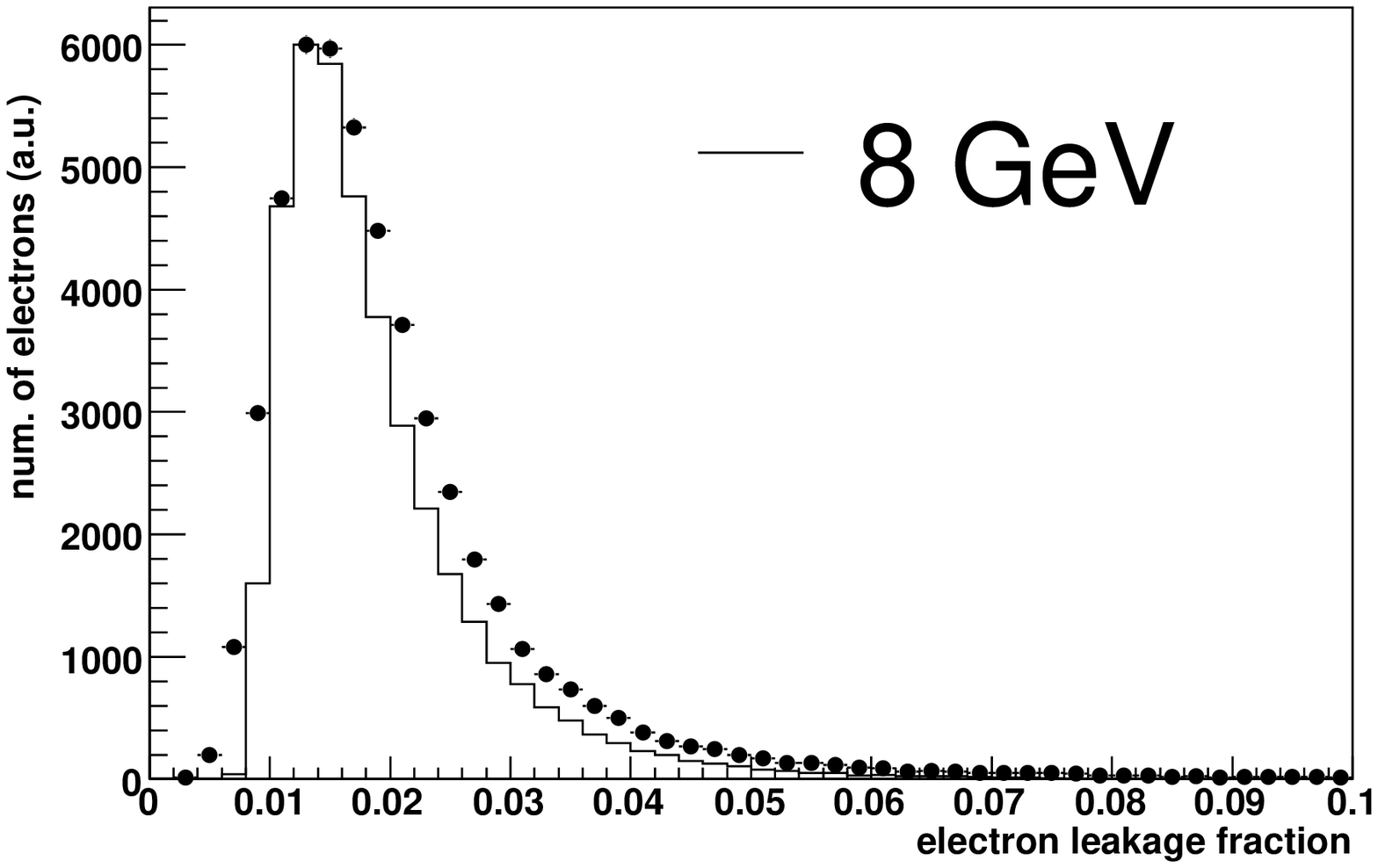}
\end{minipage}
\begin{minipage}{0.475\textwidth}
\includegraphics[width=2.7in]{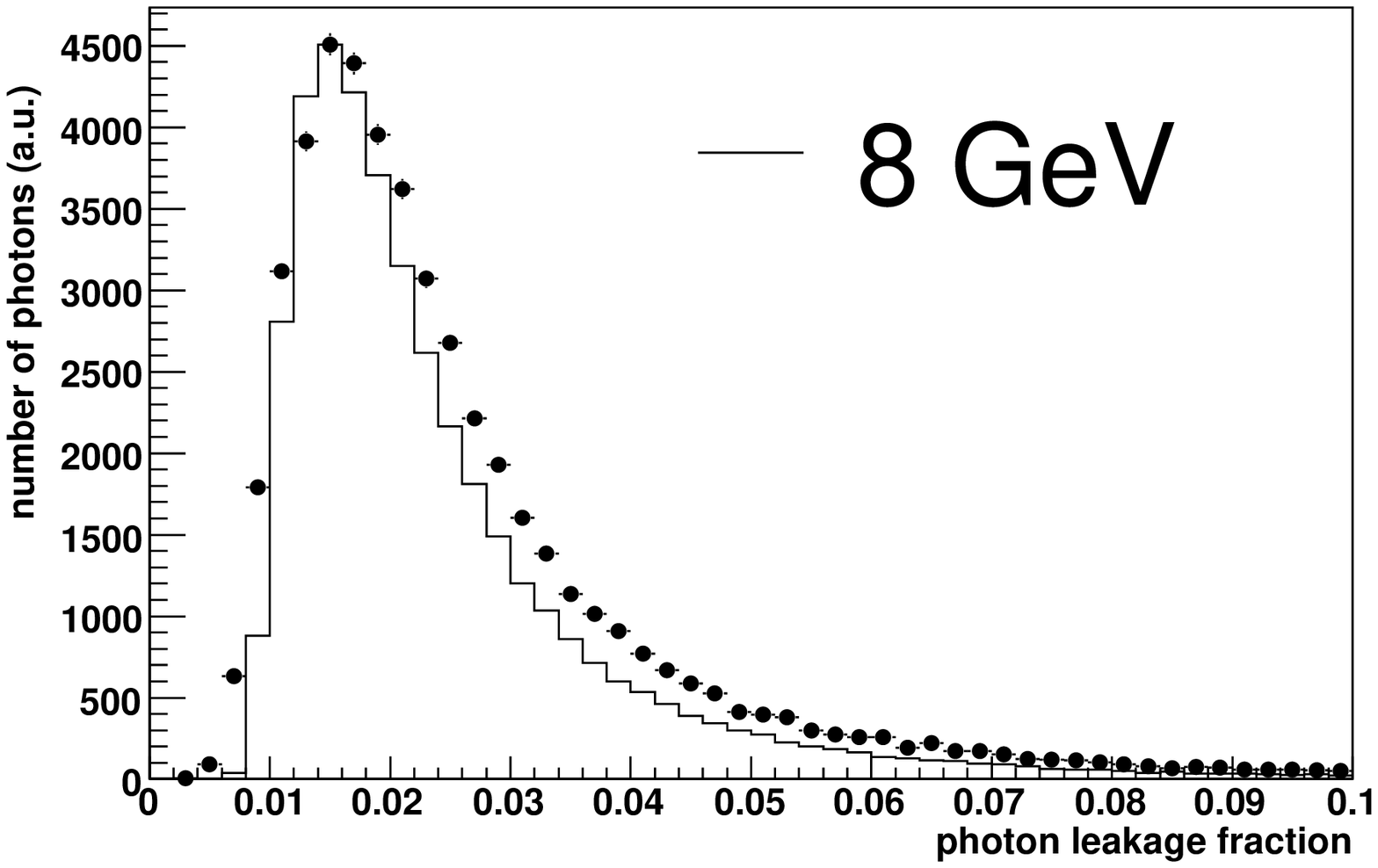}
\end{minipage}
\end{center}
\caption{Comparisons of the electron (left) and photon (right) energy leakage fraction $f_l$ for incident energies of 25 GeV (top) to 8 GeV
 (bottom), showing the {\sc geant4} calculation (points) and our parameterization (histogram) from Eqn.~\ref{leakageModelEqn}, showing some disagreement at low
 energy. The final parameterization shown in Figs.~\ref{fig:compareGeant2}-\ref{fig:compareGeant3} includes the ad-hoc correction to Eqn.~\ref{leakageModelEqn} discussed in Sec.~\ref{sec:electronLeakage}. 
}
\label{fig:compareGeantUntuned1}
\end{figure*}

\begin{figure*}
\begin{center}
\begin{minipage}{0.475\textwidth}
\includegraphics[width=2.7in]{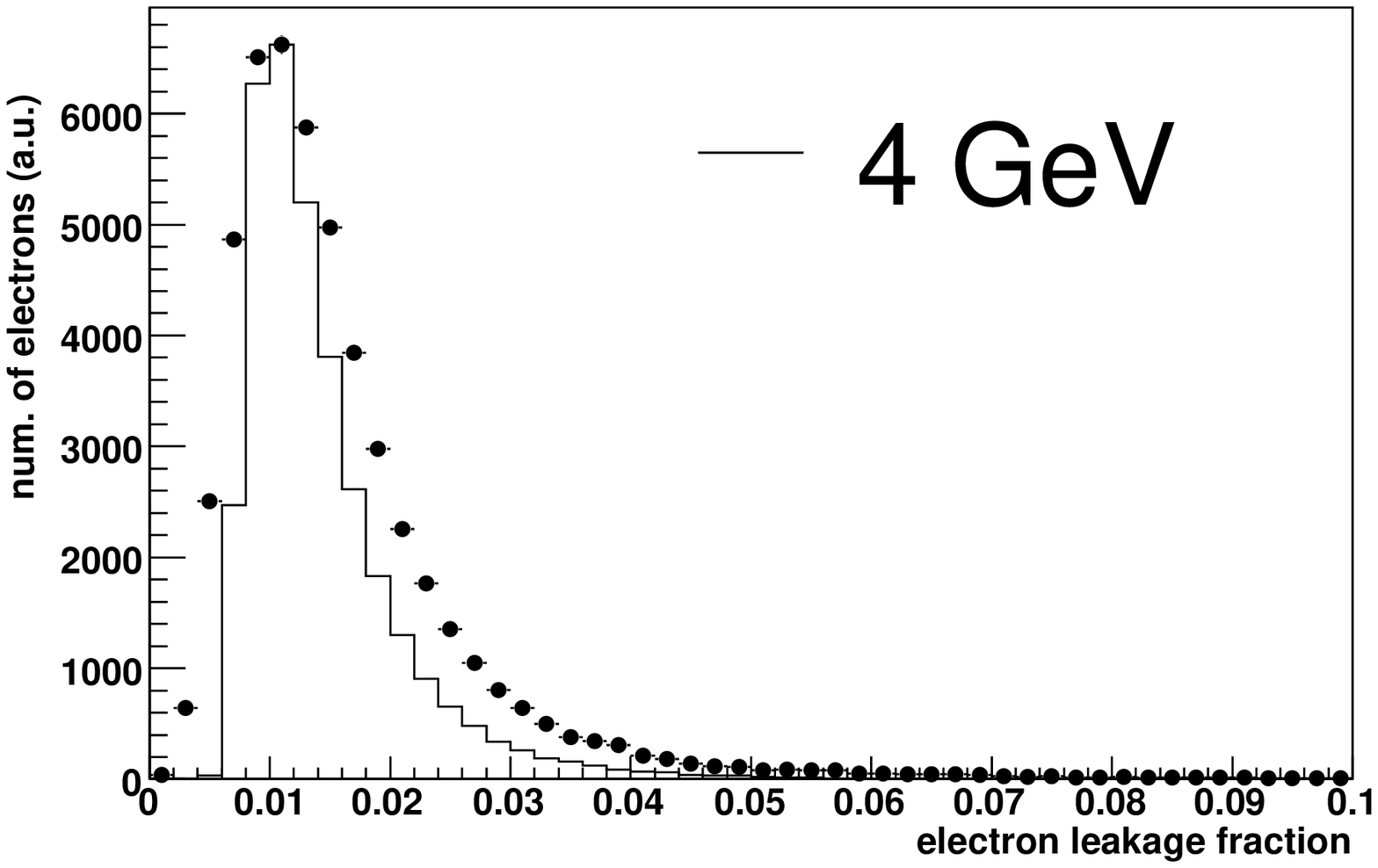}
\end{minipage}
\begin{minipage}{0.475\textwidth}
\includegraphics[width=2.7in]{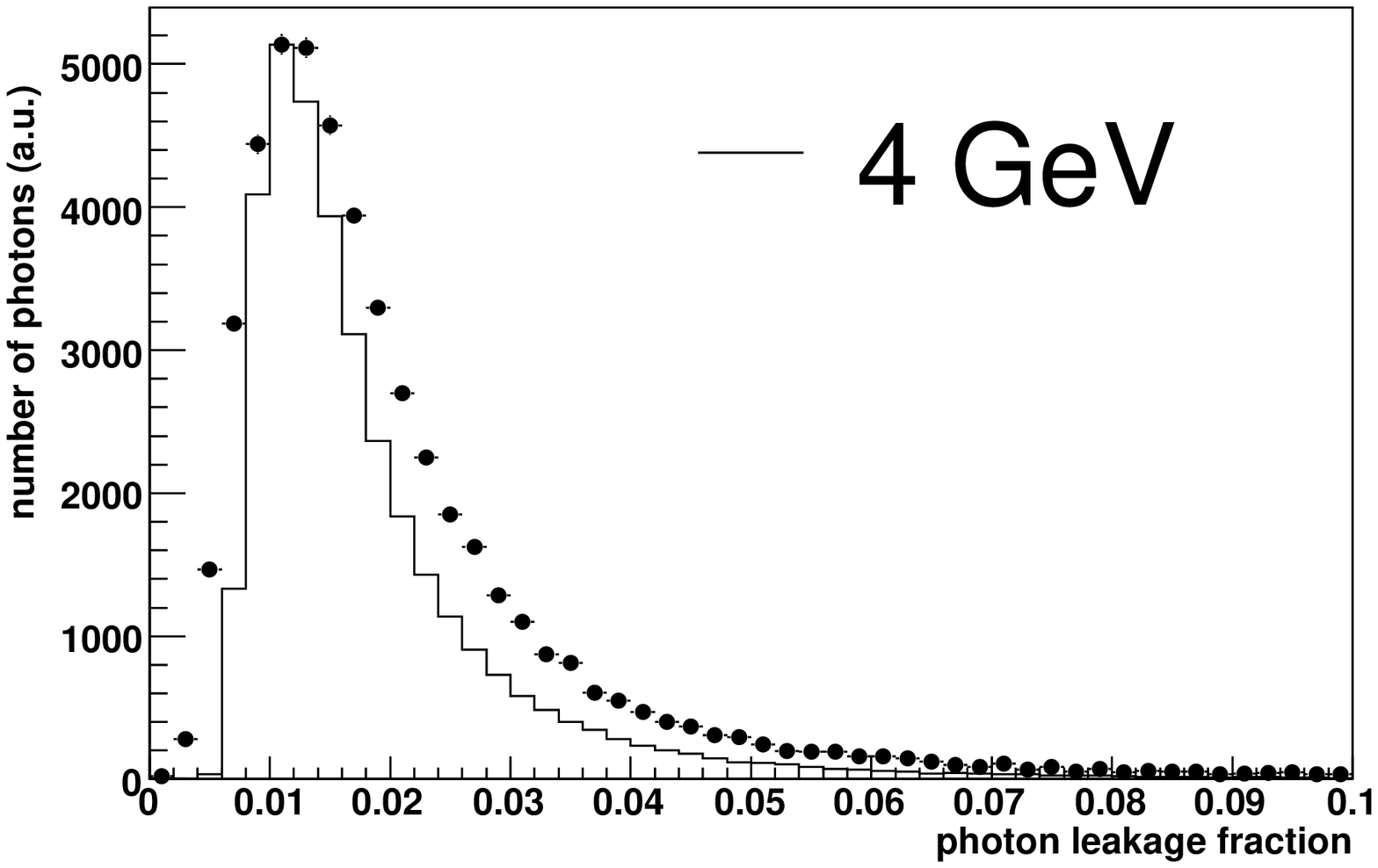}
\end{minipage}
\begin{minipage}{0.475\textwidth}
\includegraphics[width=2.7in]{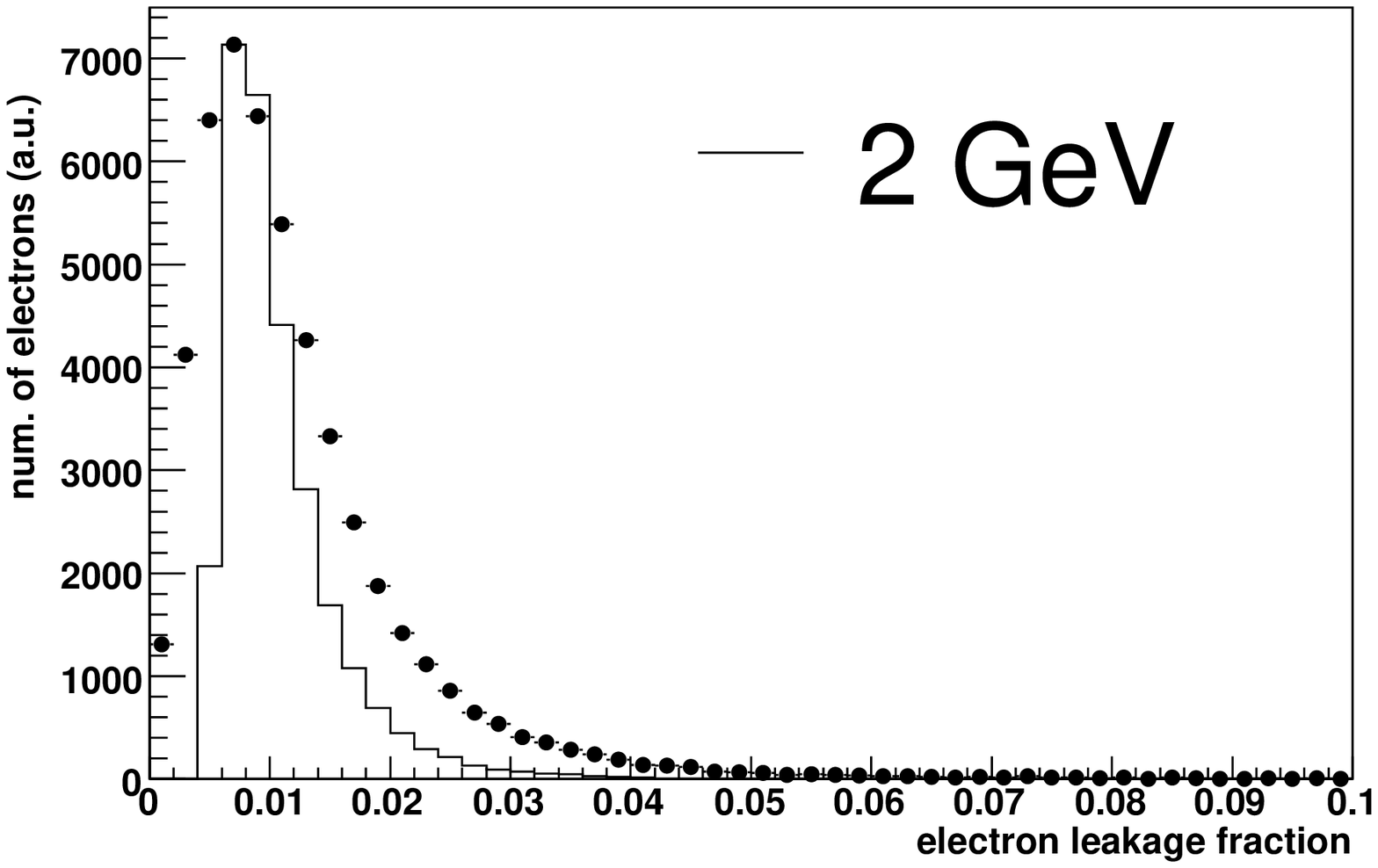}
\end{minipage}
\begin{minipage}{0.475\textwidth}
\includegraphics[width=2.7in]{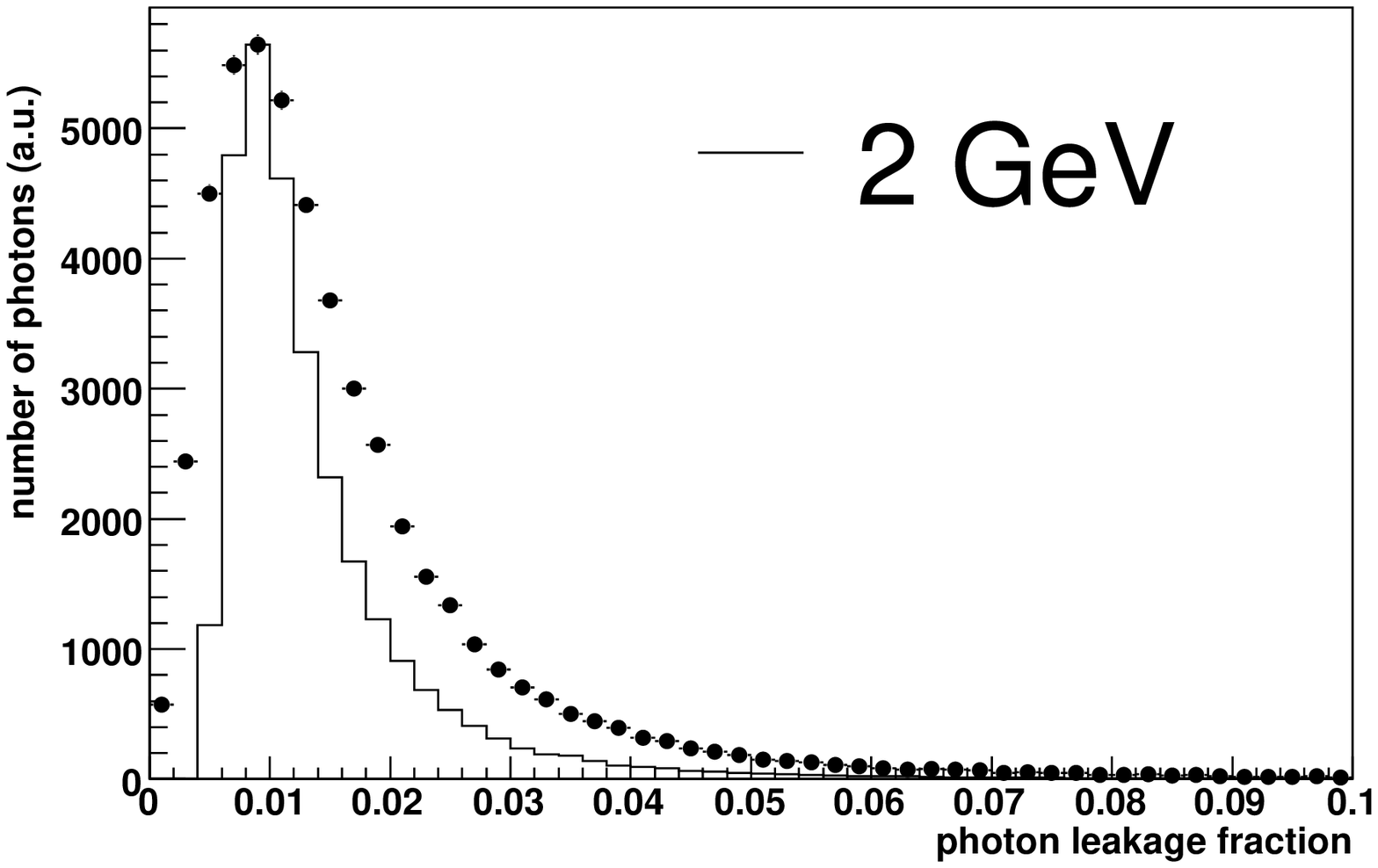}
\end{minipage}
\begin{minipage}{0.475\textwidth}
\includegraphics[width=2.7in]{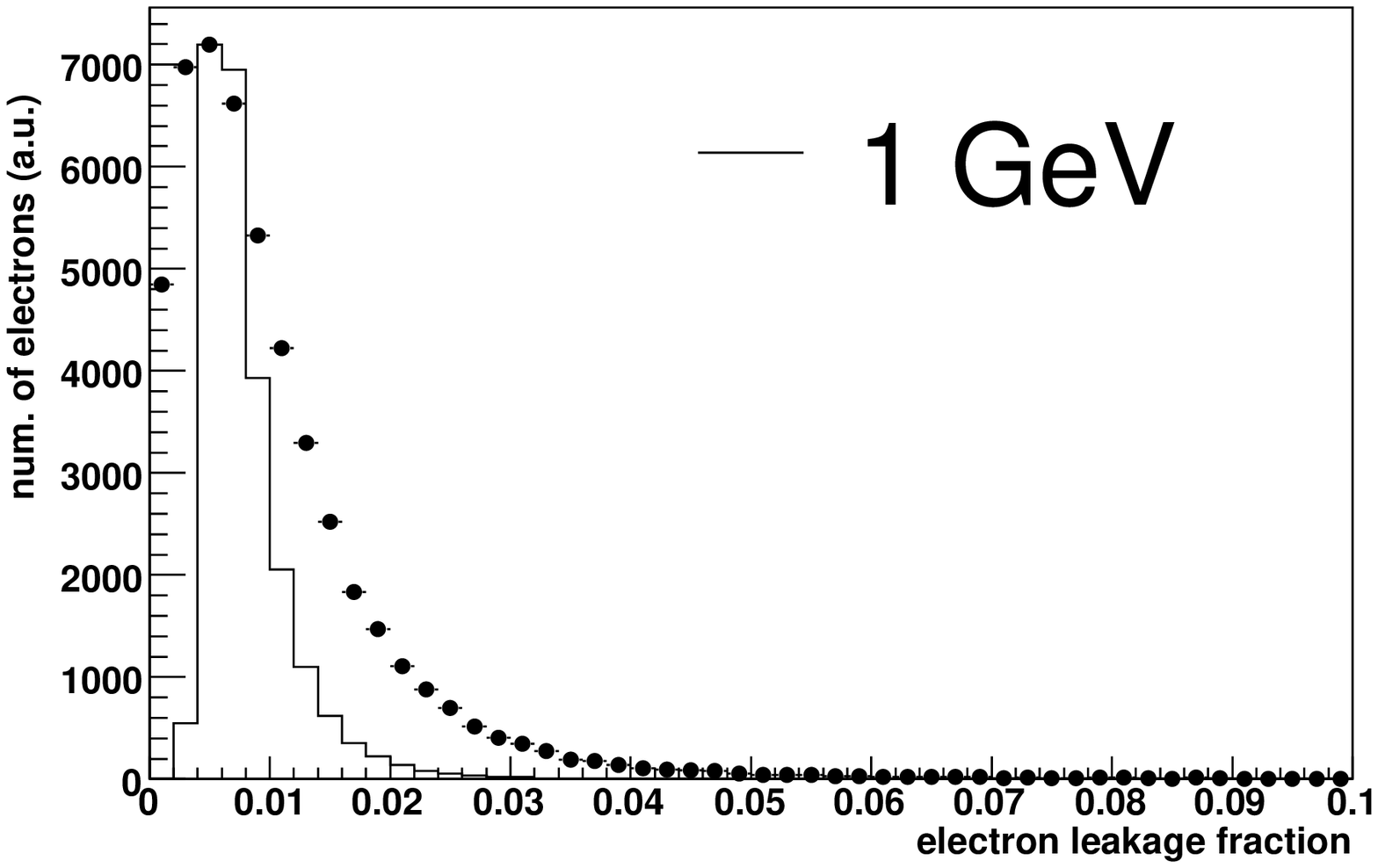}
\end{minipage}
\begin{minipage}{0.475\textwidth}
\includegraphics[width=2.7in]{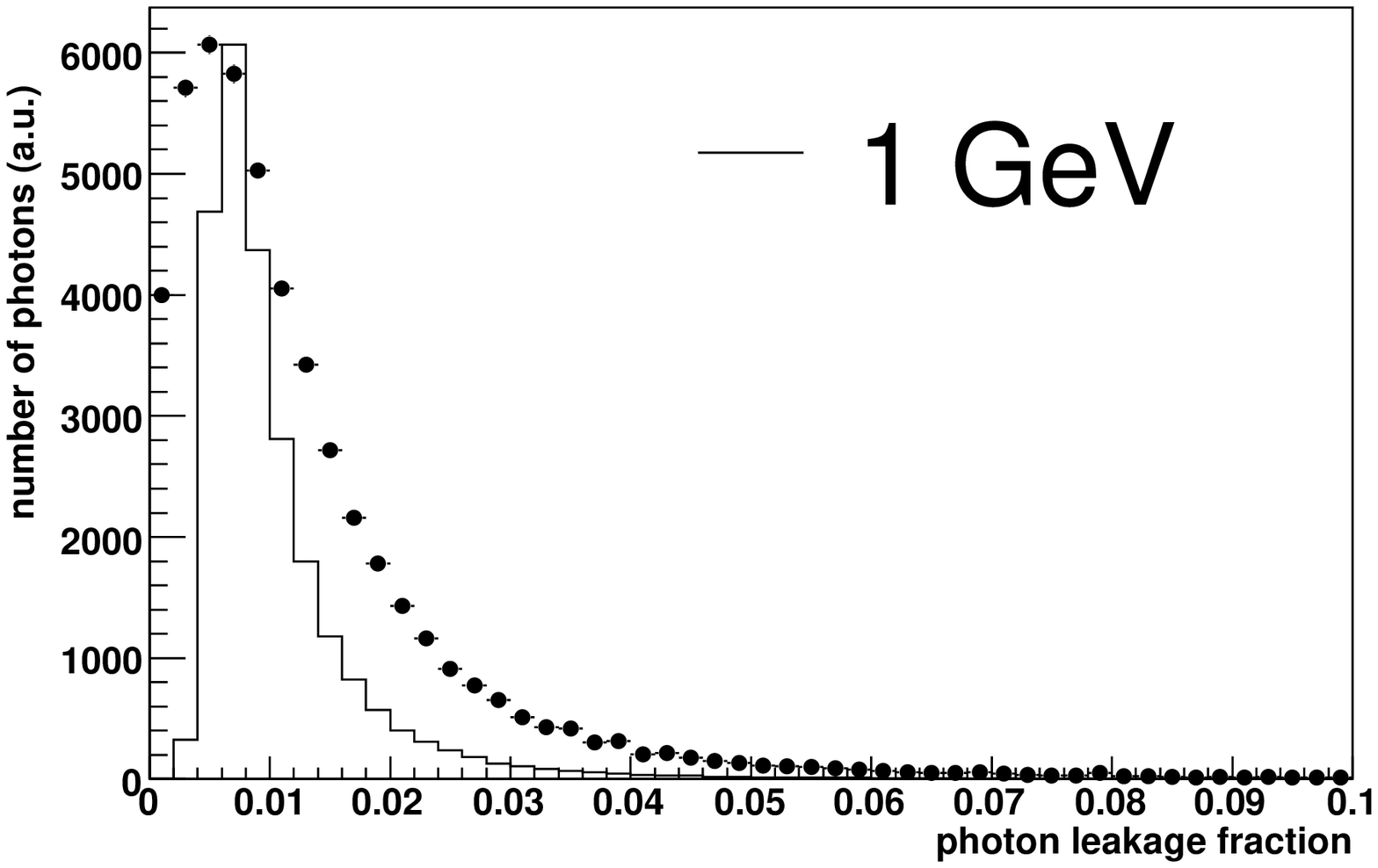}
\end{minipage}
\end{center}
\caption{Comparisons of the electron (left) and photon (right) energy leakage fraction $f_l$ for incident energies of 4 GeV (top) to 1 GeV
 (bottom), showing the {\sc geant4} calculation  (points) and the parameterization (histogram) of Eqn.~\ref{leakageModelEqn}, showing increasing disagreement at lower 
 energy. The final parameterization shown in Figs.~\ref{fig:compareGeant2}-\ref{fig:compareGeant3} includes the ad-hoc correction to Eqn.~\ref{leakageModelEqn} discussed in Sec.~\ref{sec:electronLeakage}. }
\label{fig:compareGeantUntuned2}
\end{figure*}

\begin{figure*}
\begin{center}
\begin{minipage}{0.475\textwidth}
\includegraphics[width=2.7in]{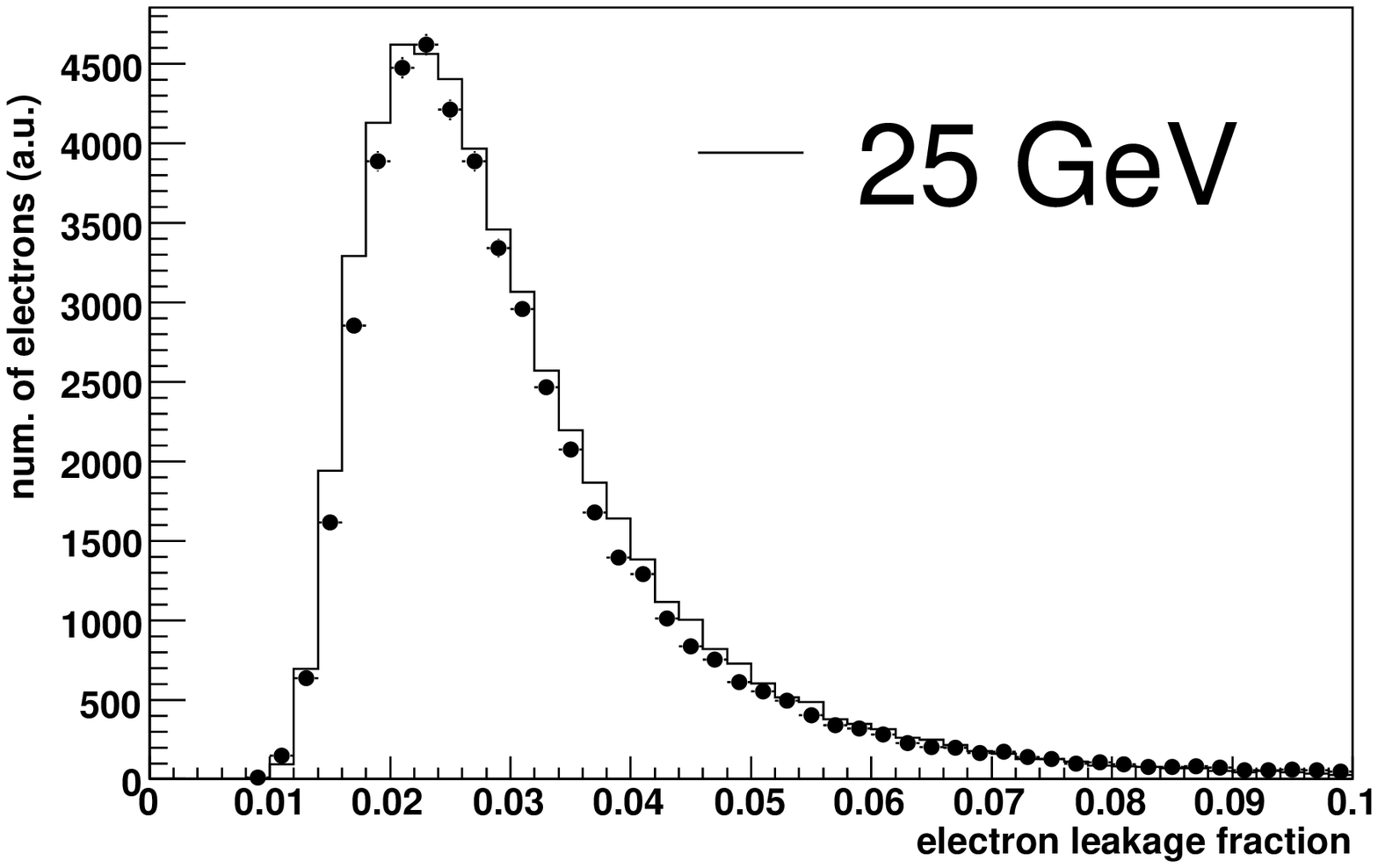}
\end{minipage}
\begin{minipage}{0.475\textwidth}
\includegraphics[width=2.7in]{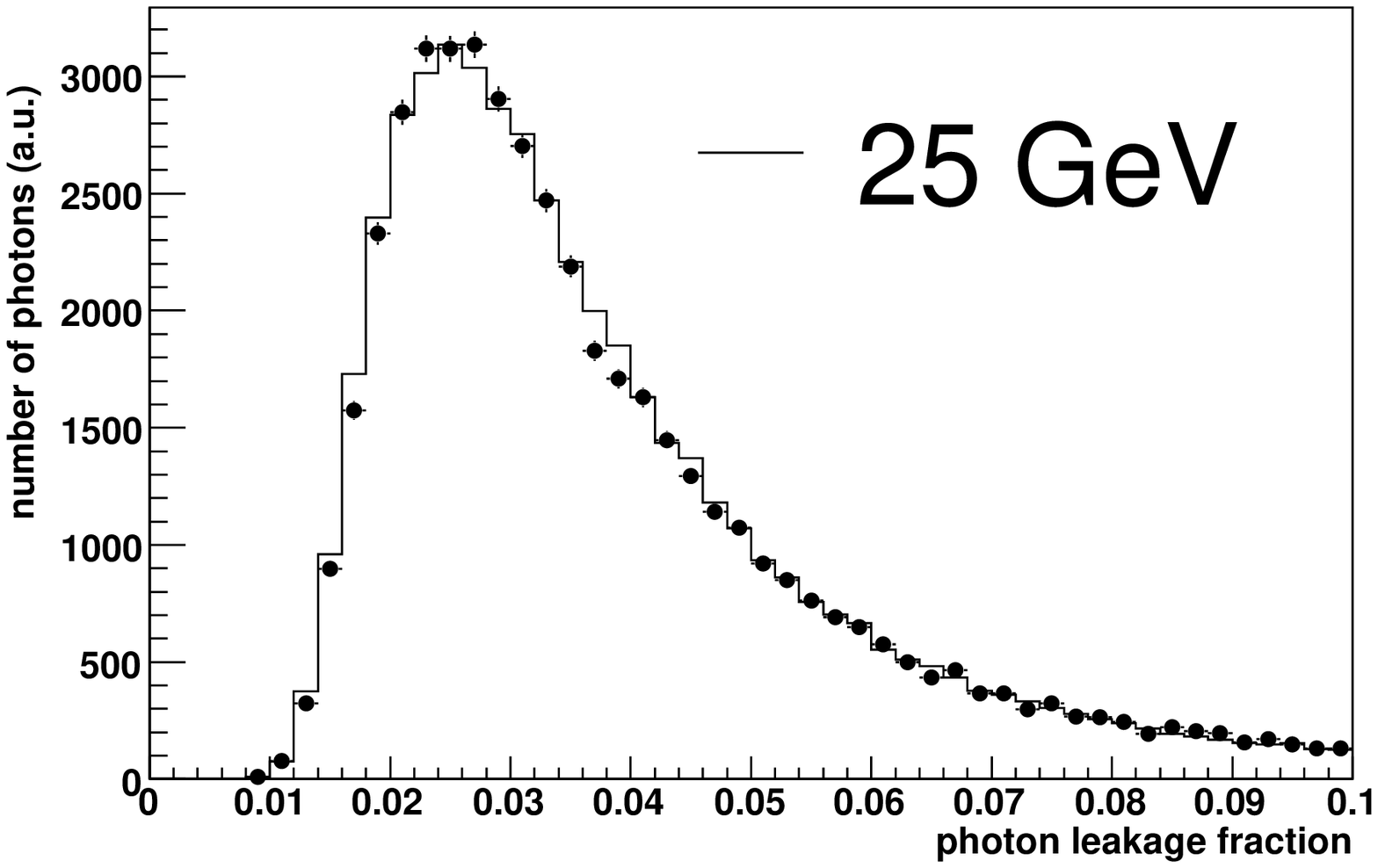}
\end{minipage}
\begin{minipage}{0.475\textwidth}
\includegraphics[width=2.7in]{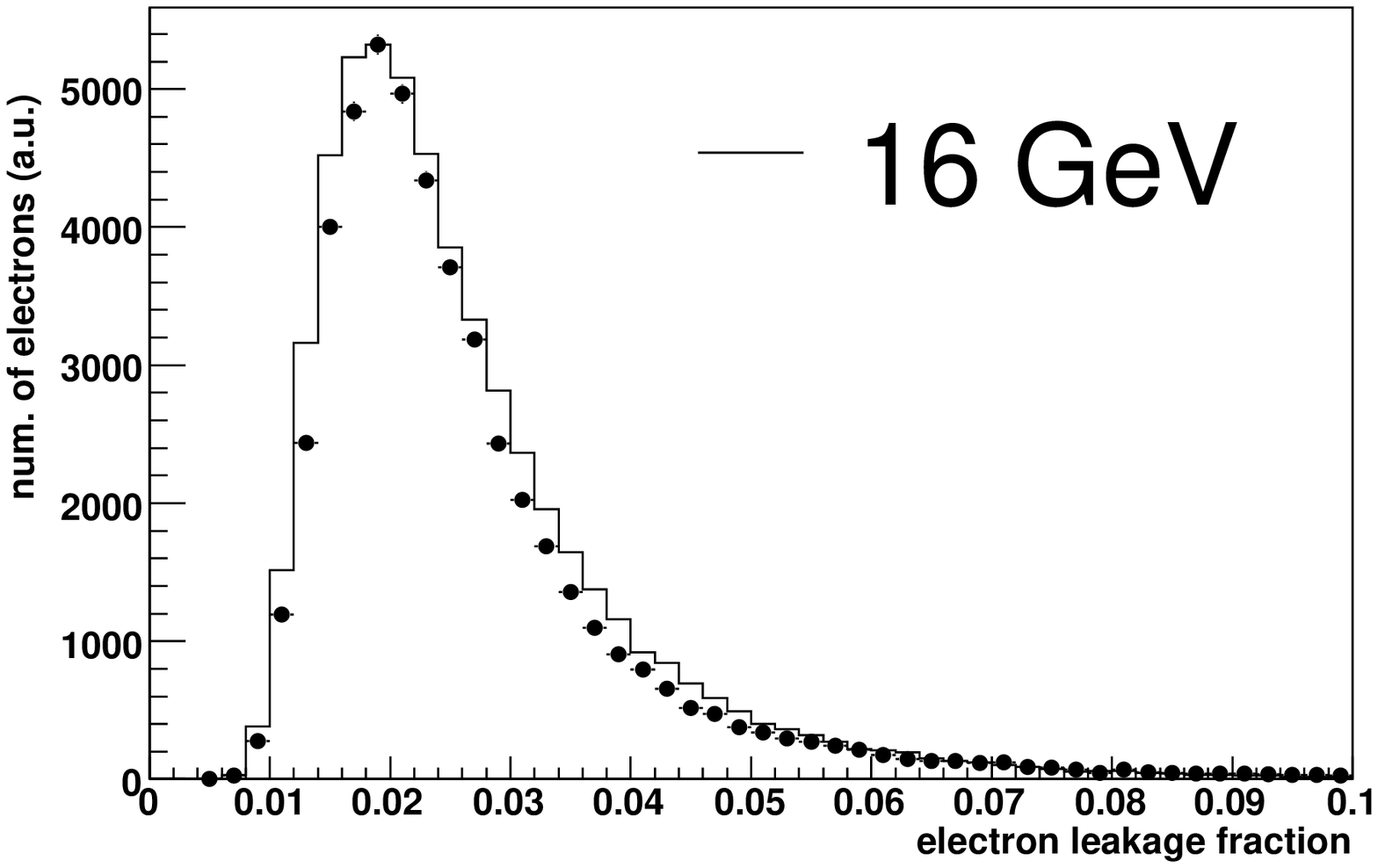}
\end{minipage}
\begin{minipage}{0.475\textwidth}
\includegraphics[width=2.7in]{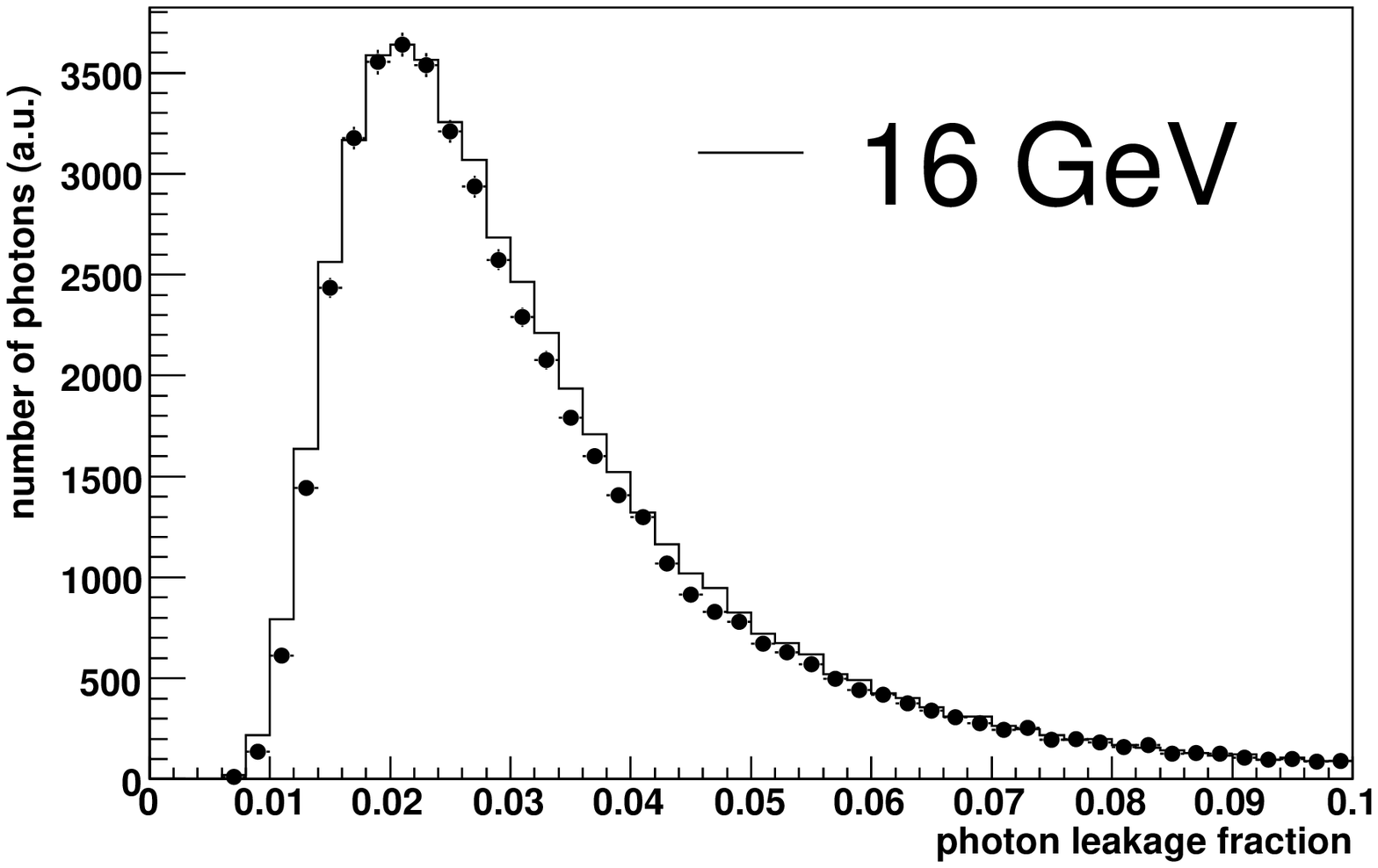}
\end{minipage}
\begin{minipage}{0.475\textwidth}
\includegraphics[width=2.7in]{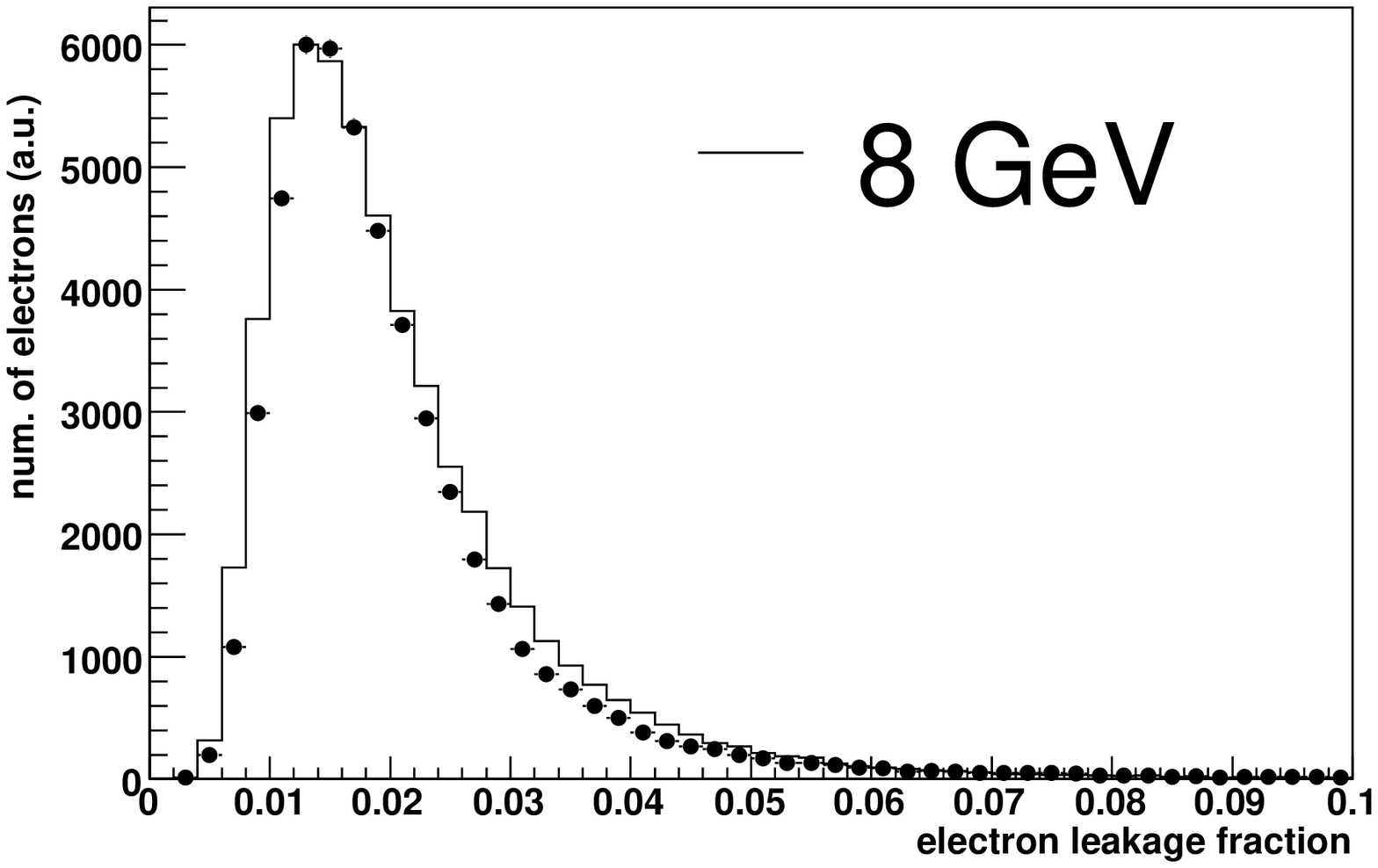}
\end{minipage}
\begin{minipage}{0.475\textwidth}
\includegraphics[width=2.7in]{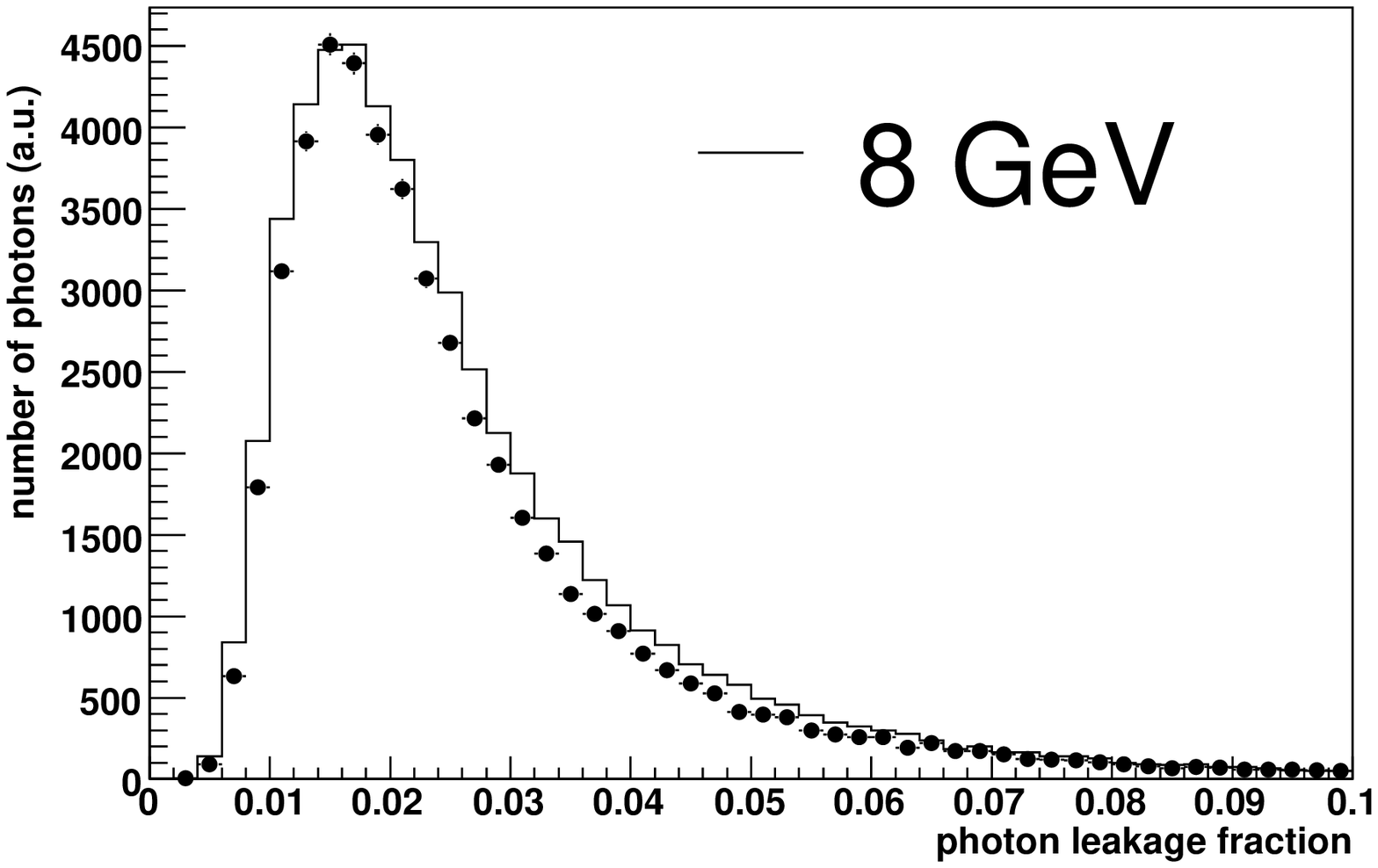}
\end{minipage}
\end{center}
\caption{Comparisons of the {\sc geant4} model calculation (points) and our tuned parameterization (histogram)
of the distribution of the energy leakage fraction $f_l$ for electrons (left) and photons 
(right) with energies ranging from of 25 GeV (top) to 8 GeV (bottom).  The ad-hoc tuning improves the agreement at low energies. 
}
\label{fig:compareGeant2}
\end{figure*}

\begin{figure*}
\begin{center}
\begin{minipage}{0.475\textwidth}
\includegraphics[width=2.7in]{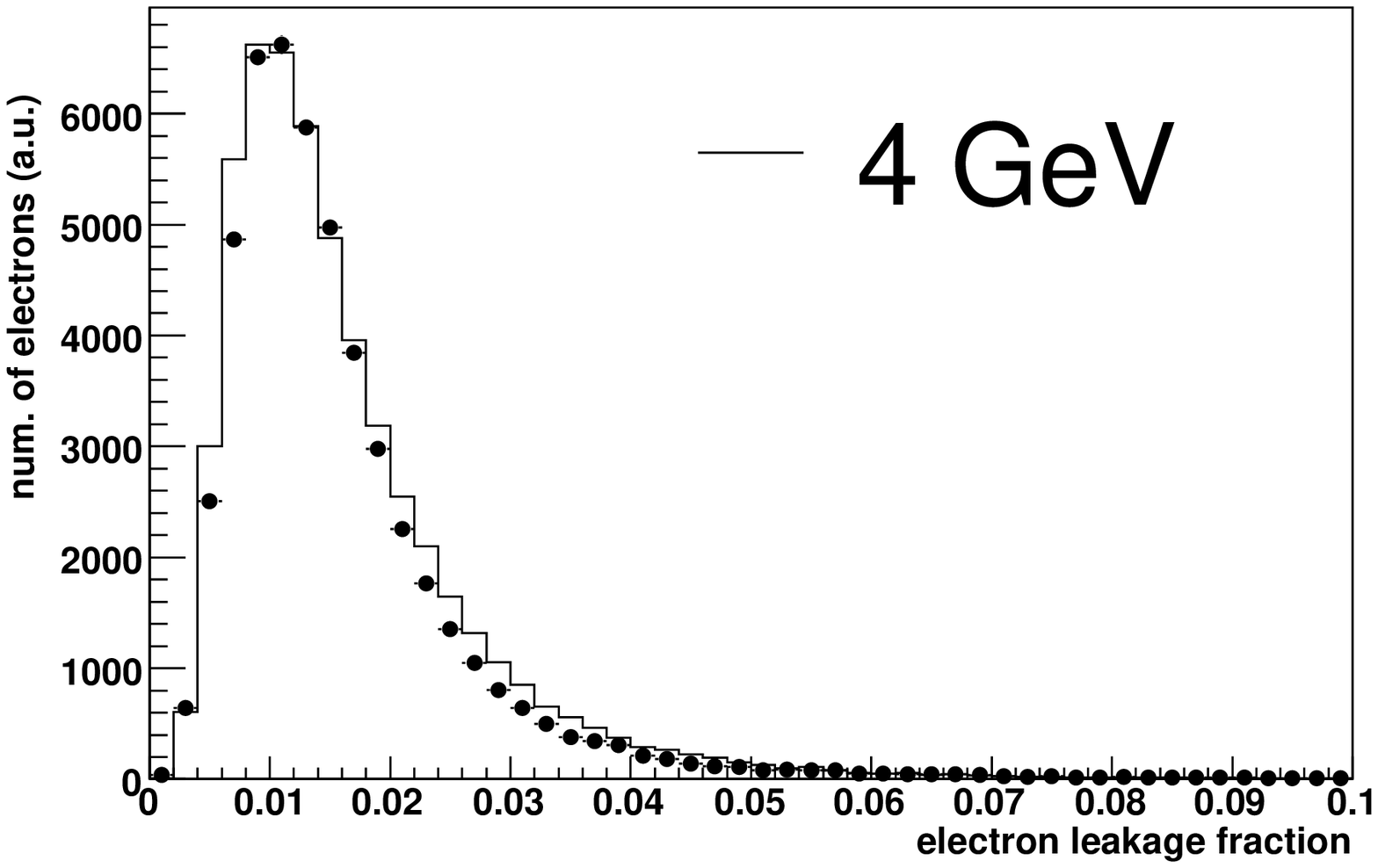}
\end{minipage}
\begin{minipage}{0.475\textwidth}
\includegraphics[width=2.7in]{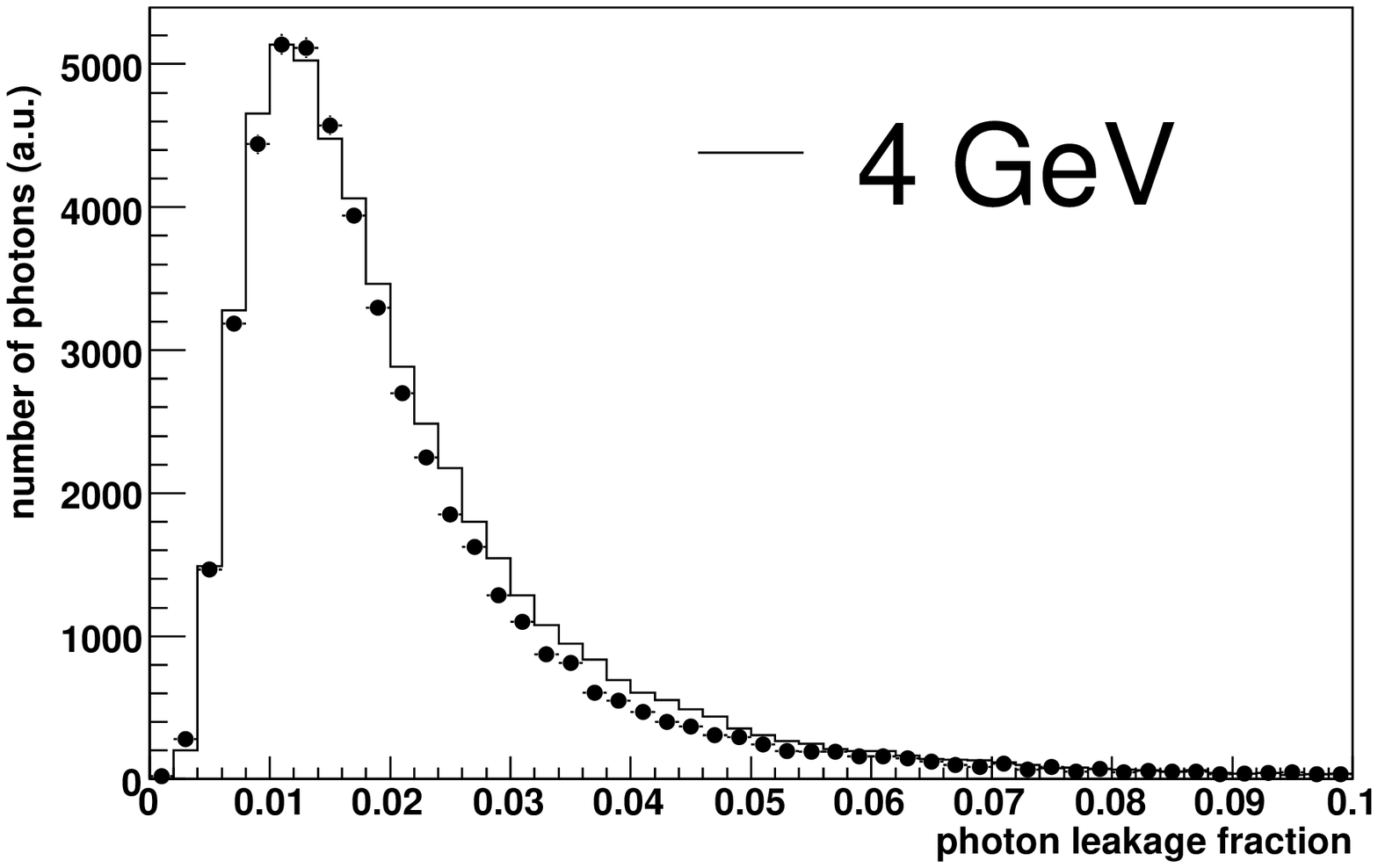}
\end{minipage}
\begin{minipage}{0.475\textwidth}
\includegraphics[width=2.7in]{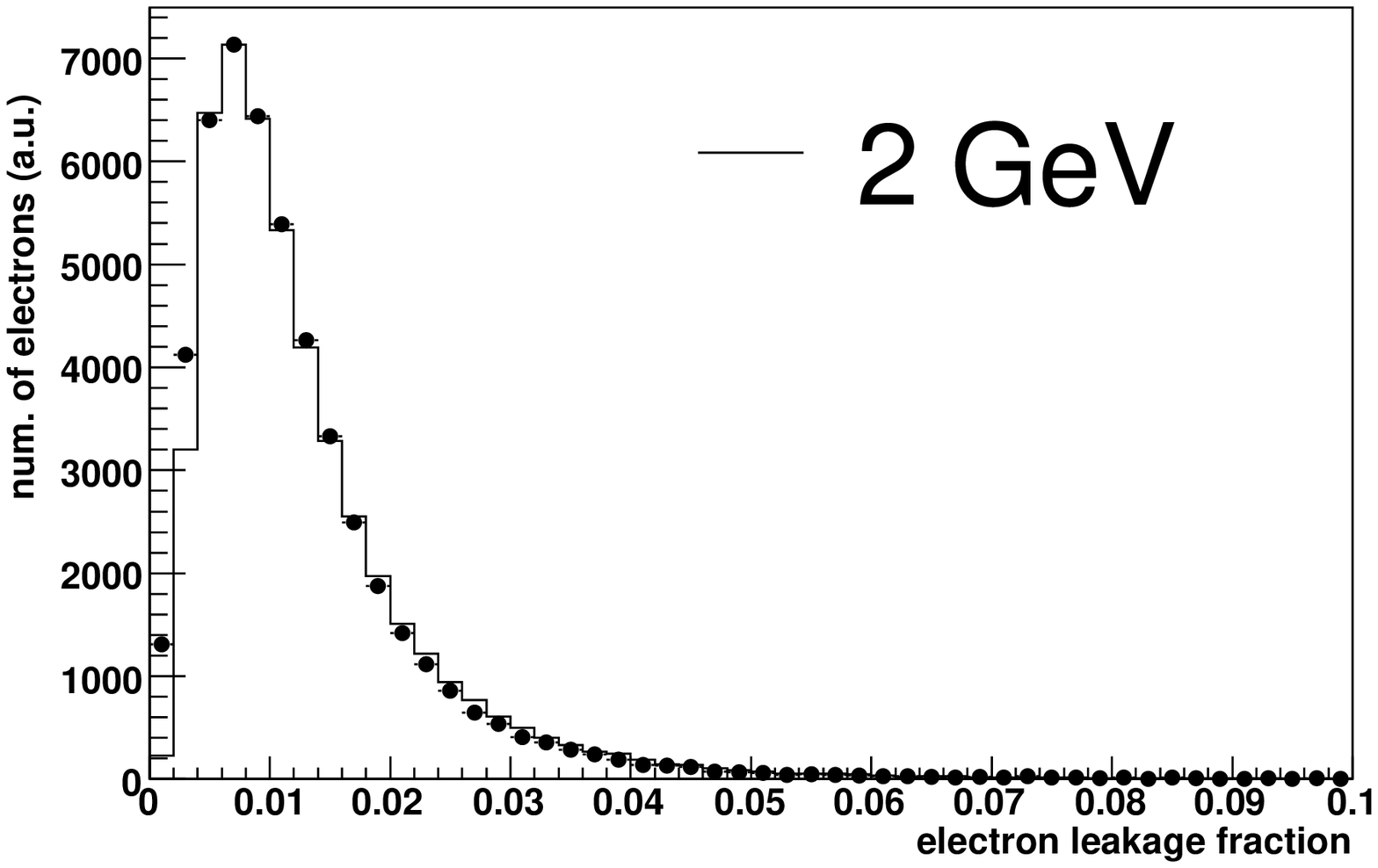}
\end{minipage}
\begin{minipage}{0.475\textwidth}
\includegraphics[width=2.7in]{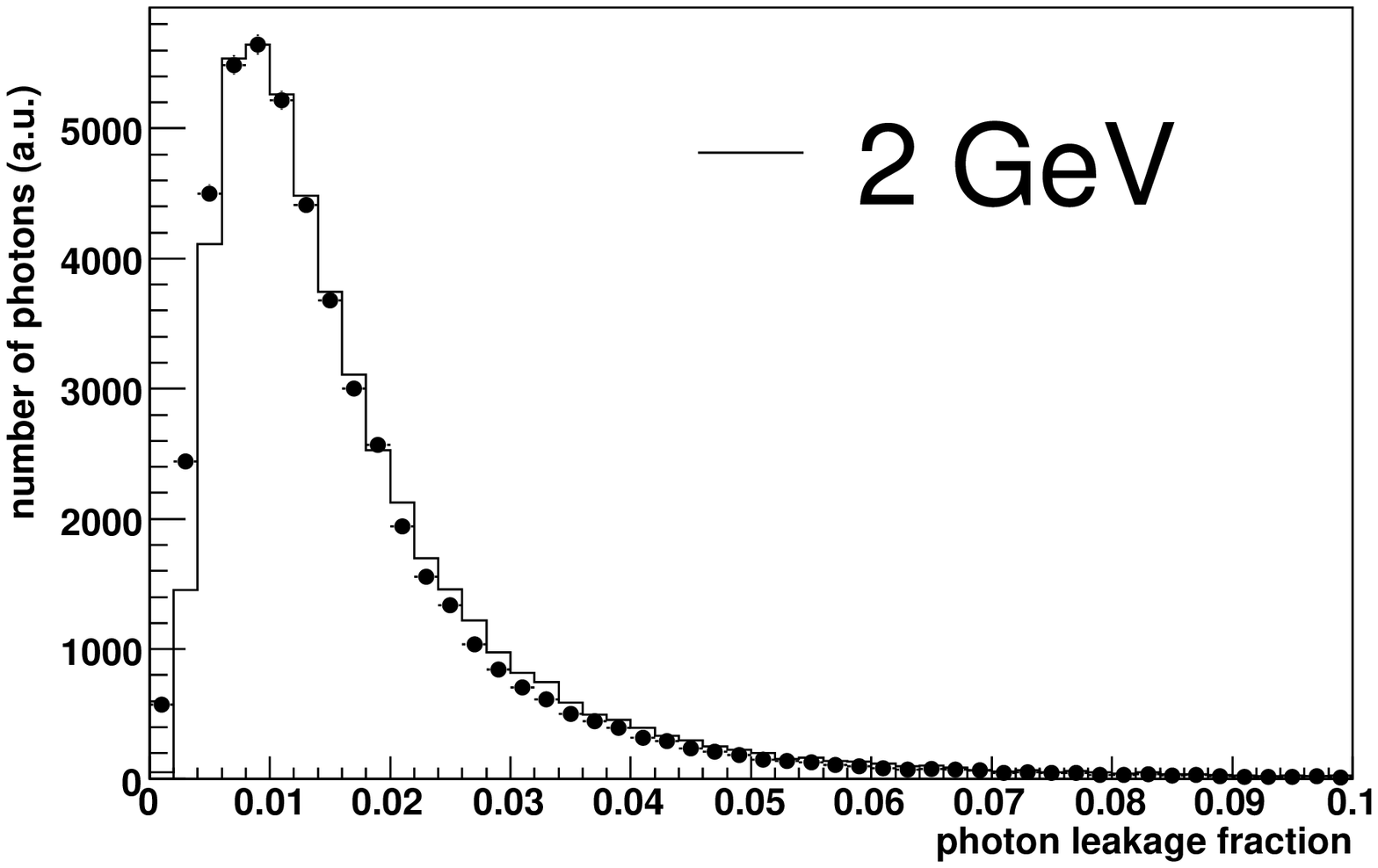}
\end{minipage}
\begin{minipage}{0.475\textwidth}
\includegraphics[width=2.7in]{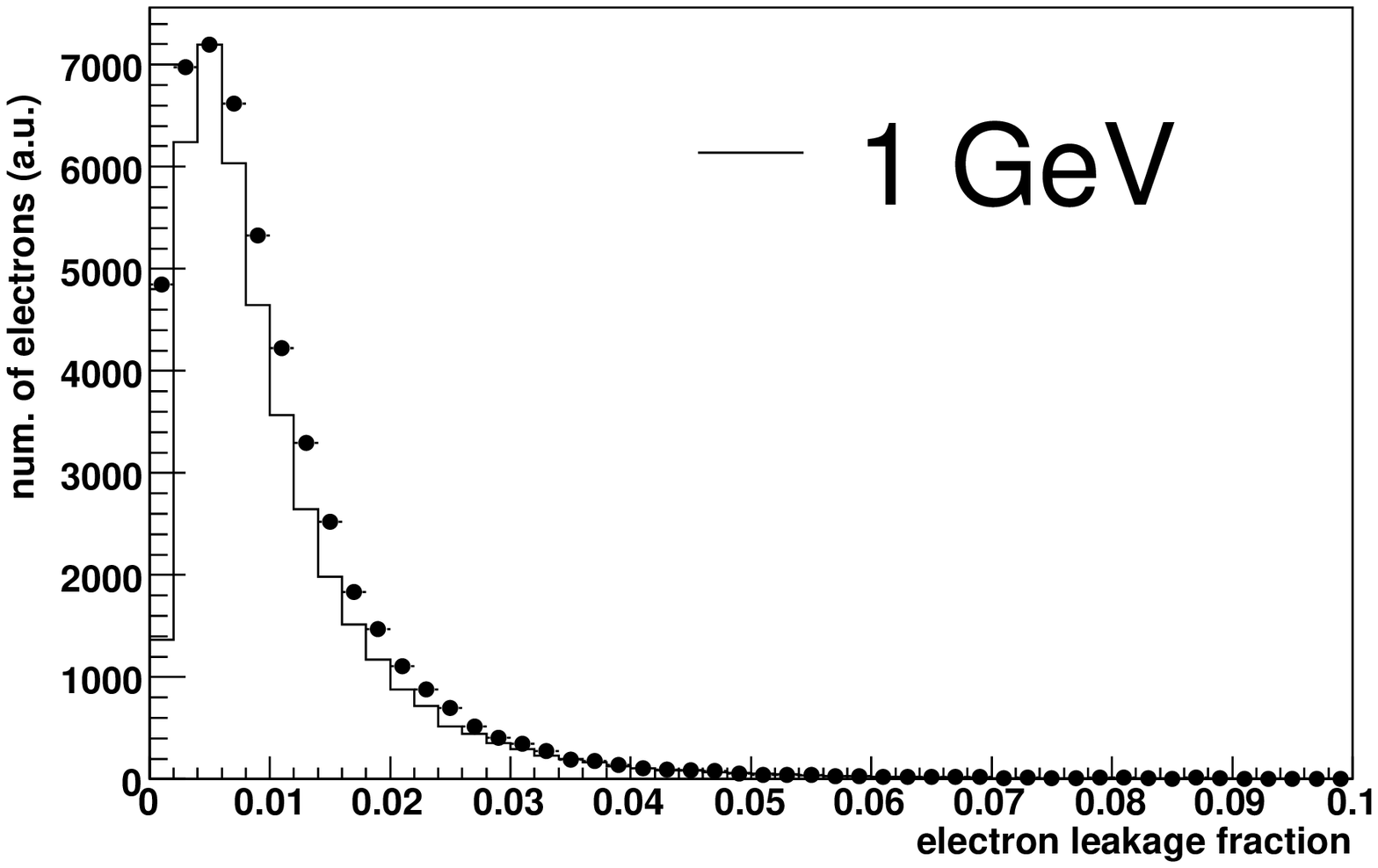}
\end{minipage}
\begin{minipage}{0.475\textwidth}
\includegraphics[width=2.7in]{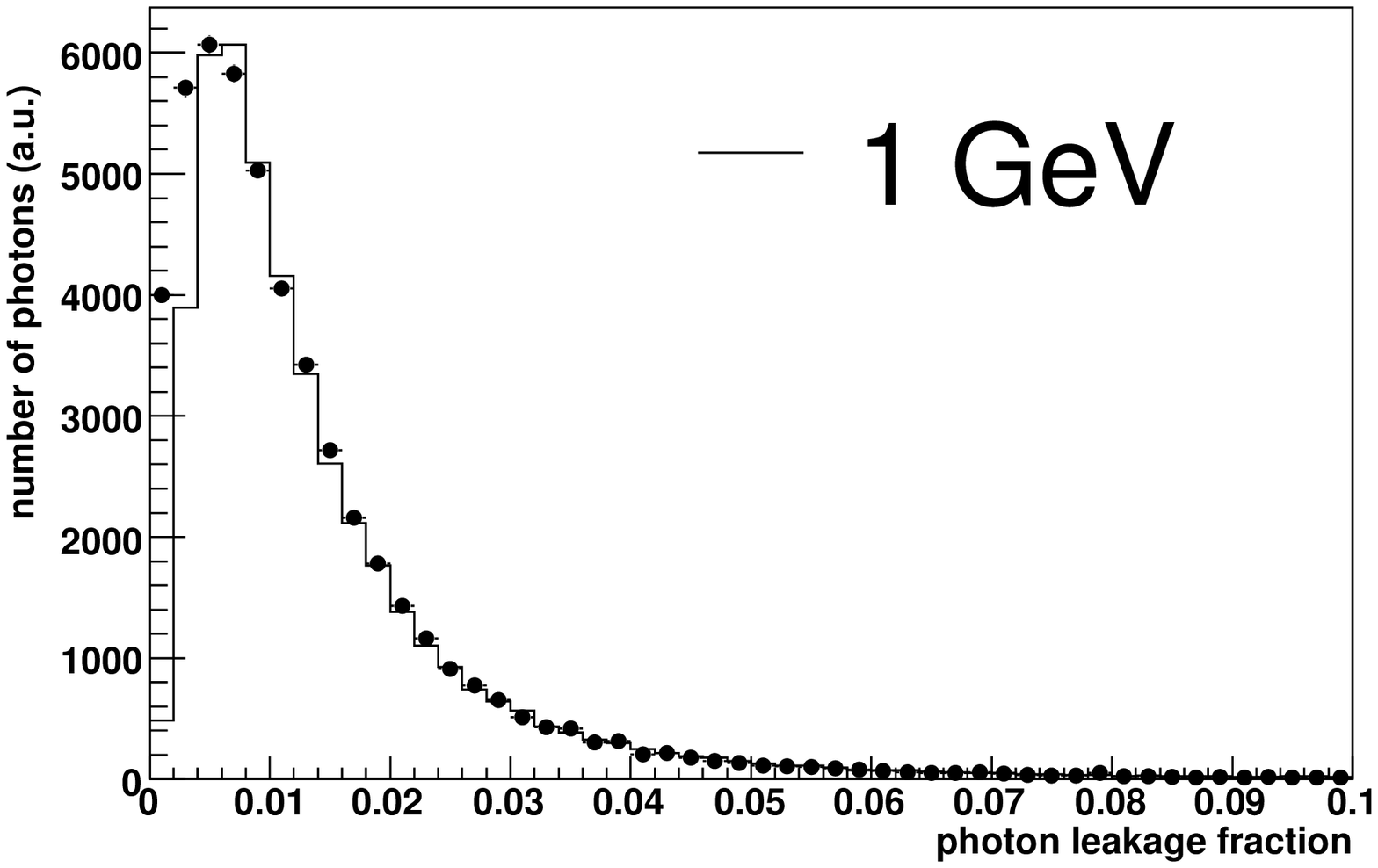}
\end{minipage}
\end{center}
\caption{Comparisons of the {\sc geant4} model calculation (points) and our tuned parameterization (histogram)
of the distribution of the energy leakage fraction $f_l$ for electrons (left) and photons 
(right) with energies ranging from of 4 GeV (top) to 1 GeV (bottom). The ad-hoc tuning improves the agreement at low energies. 
}
\label{fig:compareGeant3}
\end{figure*}

Figures~\ref{fig:compareGeantUntuned1} and~\ref{fig:compareGeantUntuned2} show the 
parameterization overlaid with the {\sc geant4} simulation for electrons with energies 
ranging from 1 to 25 GeV.  The figures show that the parameterization breaks down for 
electrons with $E < 10$ GeV.  To correct this deficiency, we smear the $\log_{10} f_l$ 
value drawn from the Gamma distribution by adding a random gaussian variable with mean 
$0.02 z$ and resolution $|0.17 z|$, where $z \equiv \log_{10} (E / 50$~GeV).  This smearing 
has a progressively larger effect on the distribution of $\log_{10} f_l$ at smaller values 
of $E$.  The combination of the Gamma distribution with this ad-hoc smearing models the 
{\sc geant4} distributions well.  Figures~\ref{fig:compareGeant2} to~\ref{fig:compareGeant3} 
show the complete parameterized model compared to the {\sc geant4} distributions.

\subsection{Photon Leakage Model}
For high energy photons the dominant interaction mechanism is electron-positron pair 
production.  We model the photon shower as a photon conversion followed by the showering 
of the resulting electron and positron.  To implement this model, we generate a random variable 
$d$ (in units of radiation lengths) representing the photon penetration depth before 
conversion.  The distribution of $d$ is given by the exponential distribution~\cite{pdg},
\begin{equation}
P(d) \propto e^{-9d/7}.
 \end{equation}
Subtracting this value of $d$ from the total thickness of the CEM in radiation lengths gives 
the remaining thickness through which the electron and positron propagate.  The energy of the 
conversion electron is obtained from the spectrum~\cite{pdg} 
\begin{equation}
\frac{d \sigma}{dy} \propto 1 - 4y/3 + 4y^2/3,
 \end{equation}
where $y$ is the fraction of the photon's energy carried by the electron.  Given the electron energy 
and the remaining calorimeter thickness, the response for the electron and positron can be 
simulated using the parameterizations given earlier, yielding the effective response for the 
original photon.  Figures~\ref{fig:compareGeant2} and~\ref{fig:compareGeant3} show that this 
model reproduces the {\sc geant4} distributions well for incident photons.

\section{Scintillator Sampling Model}
\label{sec:samplingmodel}
To first order, the scintillation light produced by the shower is proportional to the 
energy deposited in the scintillator ($E_s$).  Defining the sampling fraction as 
$f_s = E_s / [E (1-f_l)]$, where $E$ is the incident energy, the leakage fluctuations 
cause the average $f_s$ to depend on $f_l$.   In addition, there are stochastic 
fluctuations in $f_s$.  

\subsection{Leakage Dependence}
\label{sec:samplingfraction}
\begin{figure*}
\begin{center}
\includegraphics[width=2.7in]{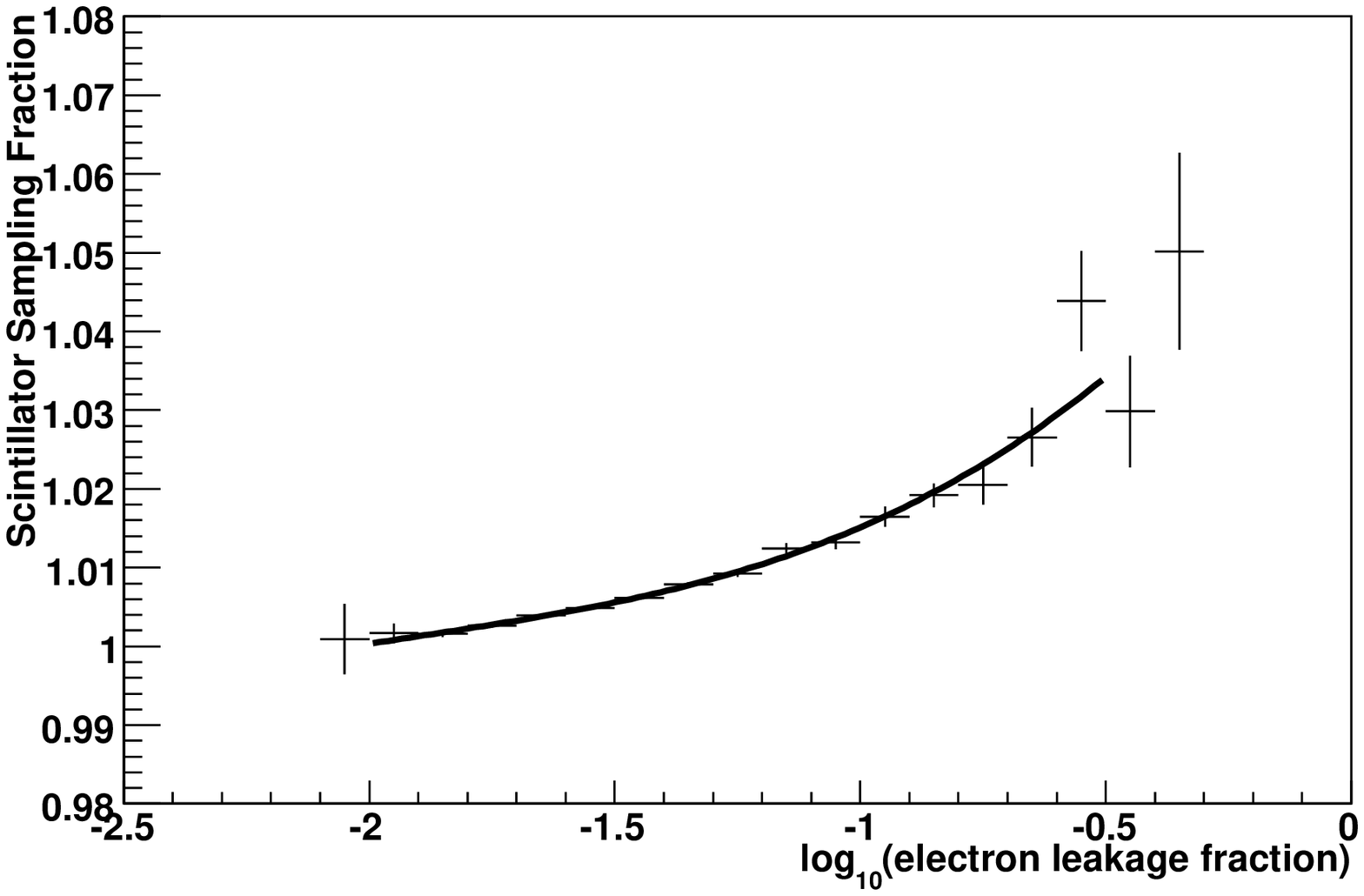}
\includegraphics[width=2.7in]{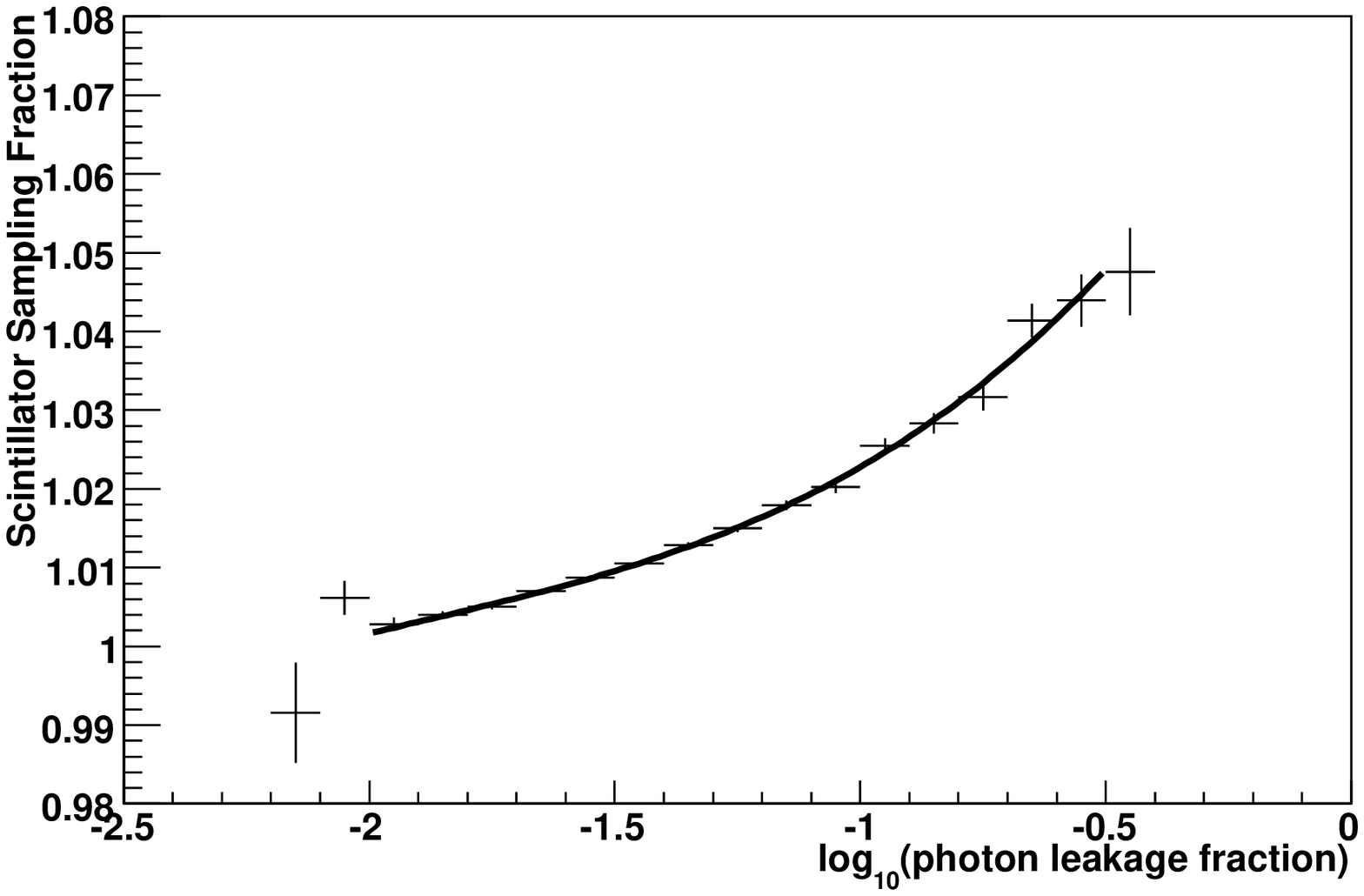}

\includegraphics[width=2.7in]{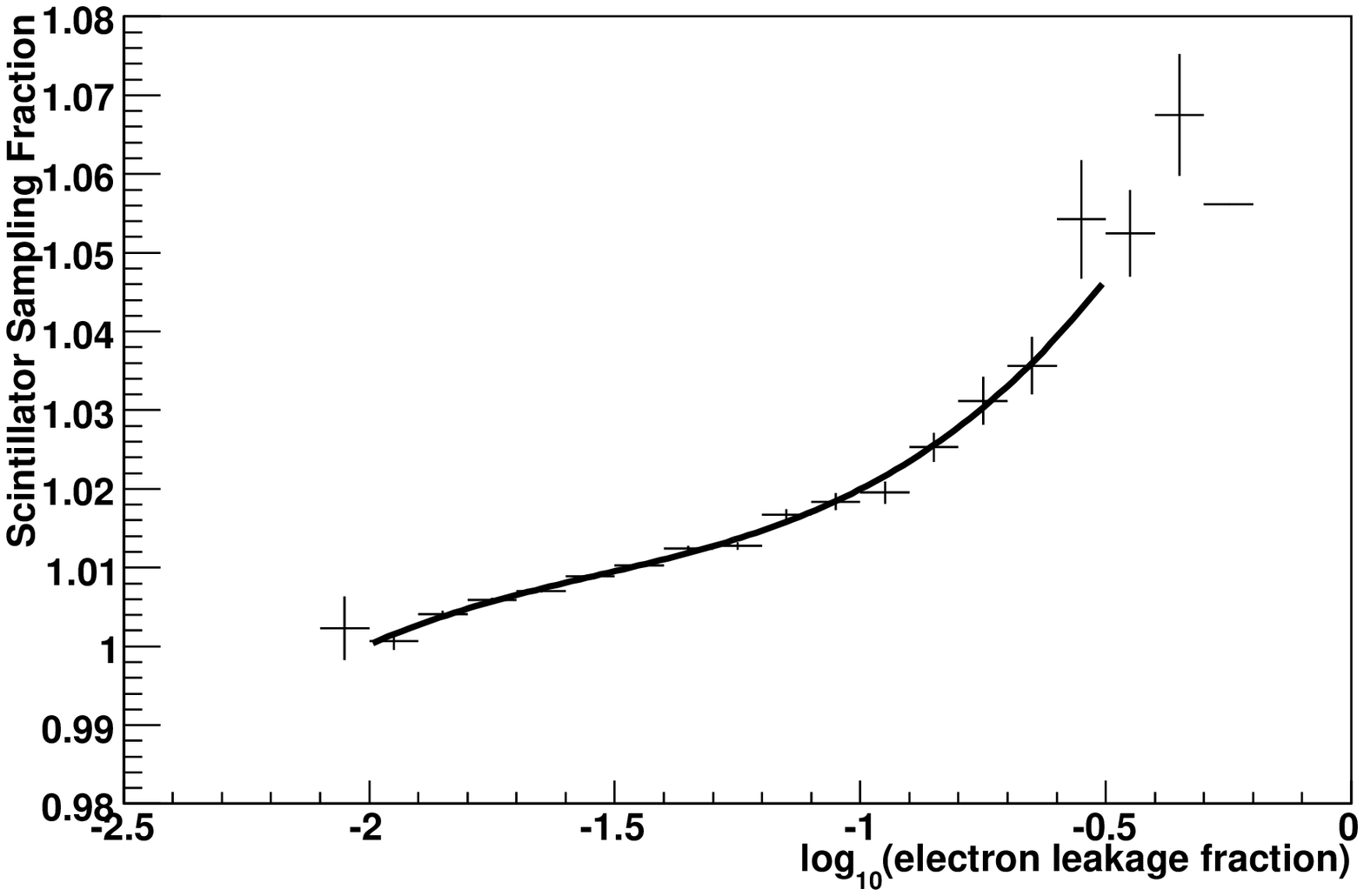}
\includegraphics[width=2.7in]{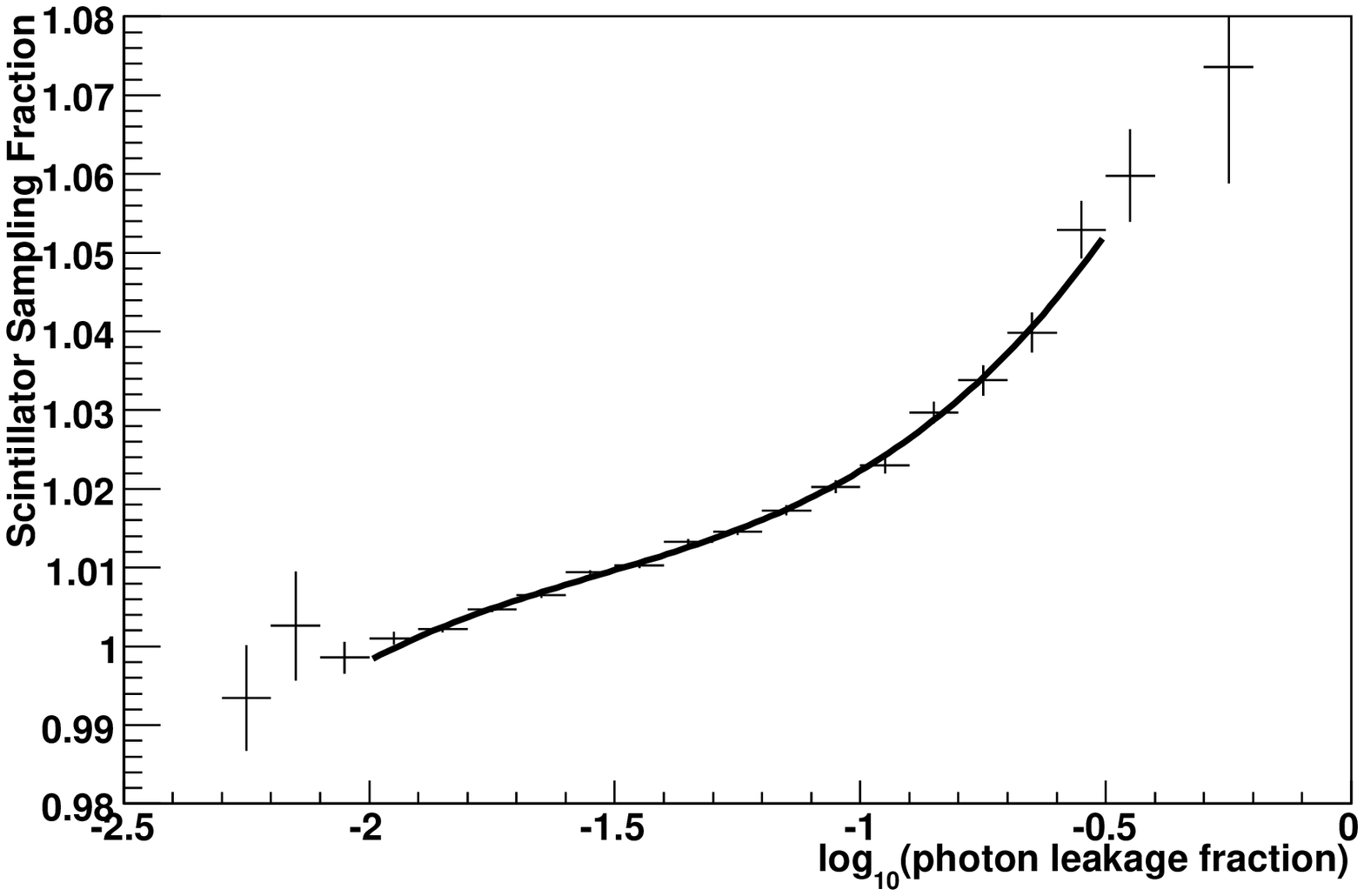}

\includegraphics[width=2.7in]{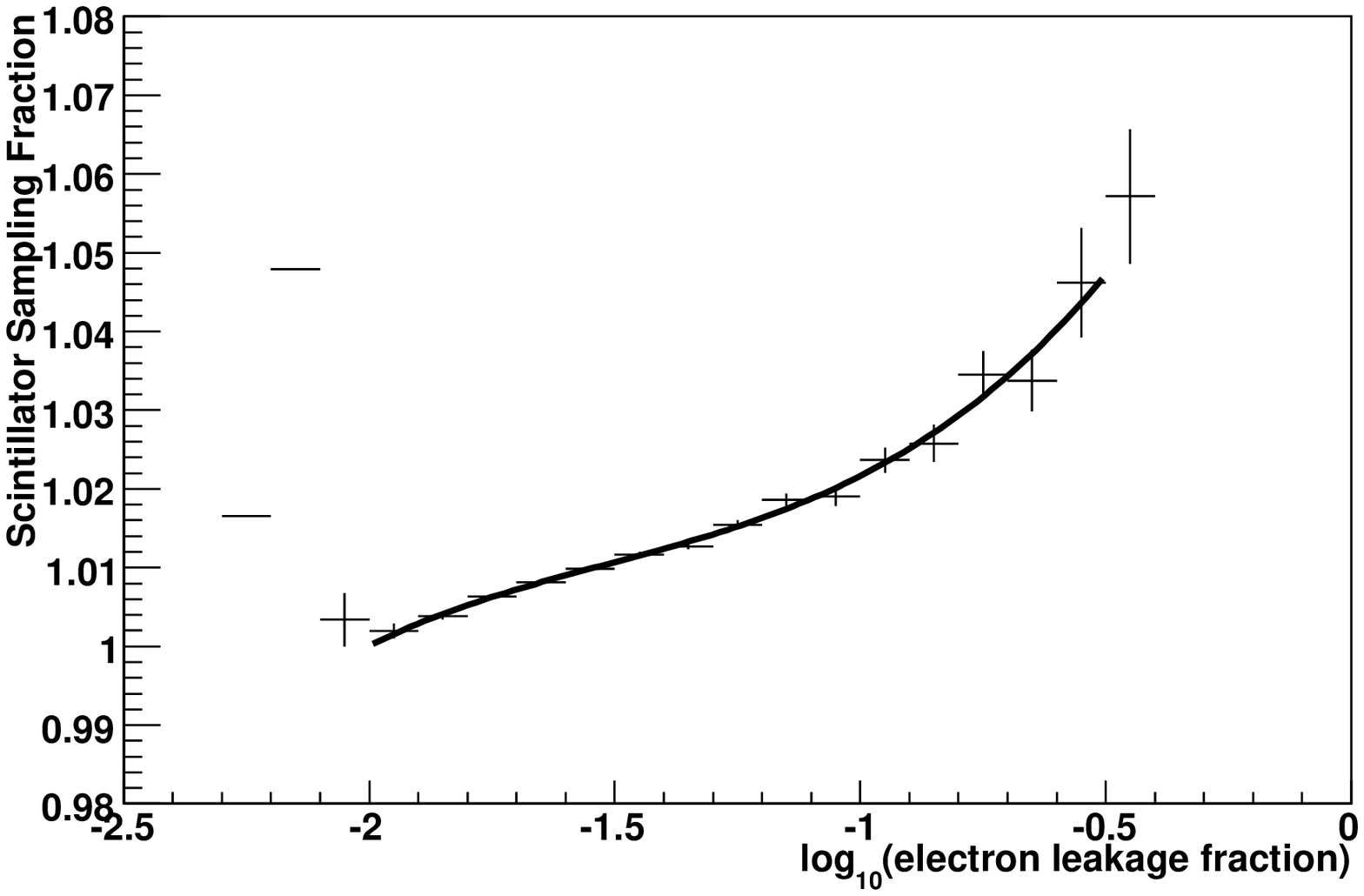}
\includegraphics[width=2.7in]{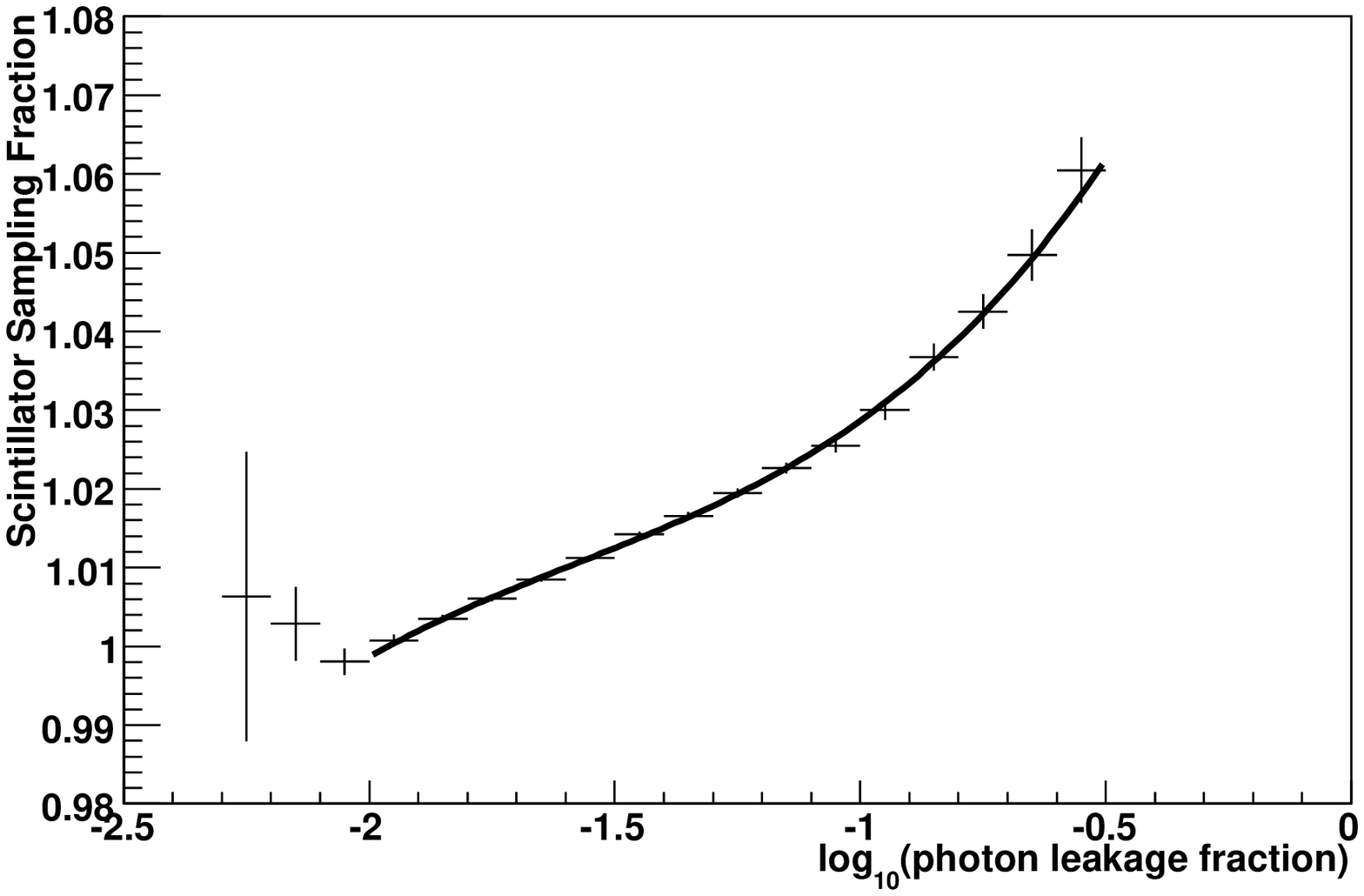}
\end{center}
\caption{ The dependence of $f_s$ on $f_l$ for electrons (left) and photons (right) 
incident on Tower 0 (top), Tower 5 (middle), or Tower 8 (bottom).  The curves are polynomial 
fits to the points calculated with the {\sc geant4} model. For each plot, $f_s$ has been rescaled such that $f_s = 1$ at $f_l = 0.01$.  }
\label{samplingGain}
\end{figure*}

We expect a positive correlation between $f_s$ and $f_l$ because showers that initiate early 
have relatively low $f_l$ and also low $f_s$, since the first $\approx 0.6$ radiation lengths 
are not instrumented and the scintillator layers follow the absorber in the subsequent layers.  
Figure~\ref{samplingGain} shows the correlation between $f_s$ and $f_l$ for electrons (photons) 
for three tower geometries:  Tower 0, Tower 5 (Tower 4) and Tower 8.  Since the absolute value 
of $f_s$ ($\approx 12$\%) is irrelevant, we rescale $f_s$ for each plot such that $f_s = 1$ at 
$f_l = 0.01$.  

Figure~\ref{samplingGain} shows that the positive correlation between $f_s$ and $f_l$ grows as 
the tower rapidity increases.  One reason for this increased correlation is the increasing 
amount of uninstrumented material in front of the calorimeter, due to oblique incidence. Since
 the upstream uninstrumented material causes the positive correlation, increasing the former
 increases the correlation. Another 
reason is the reduction in the number of scintillator layers, so that each 
 scintillator contributes
a larger fraction of the total signal. Showers with higher leakage have more signal deposited
  in the last 
scintillator, giving a larger positive correlation between $f_s$ and leakage.

The polynomial fits in Fig.~\ref{samplingGain} are used to evaluate $\langle f_s \rangle$ for 
a given value of $f_l$ in the custom simulation of electron and photon showers used for the $W$ 
boson mass measurement at CDF~\cite{cdfwmass}.  

\subsection{Sampling Resolution}
\label{sec:samplingresolution}
\begin{figure}
\begin{center}
\includegraphics[width=3.3in]{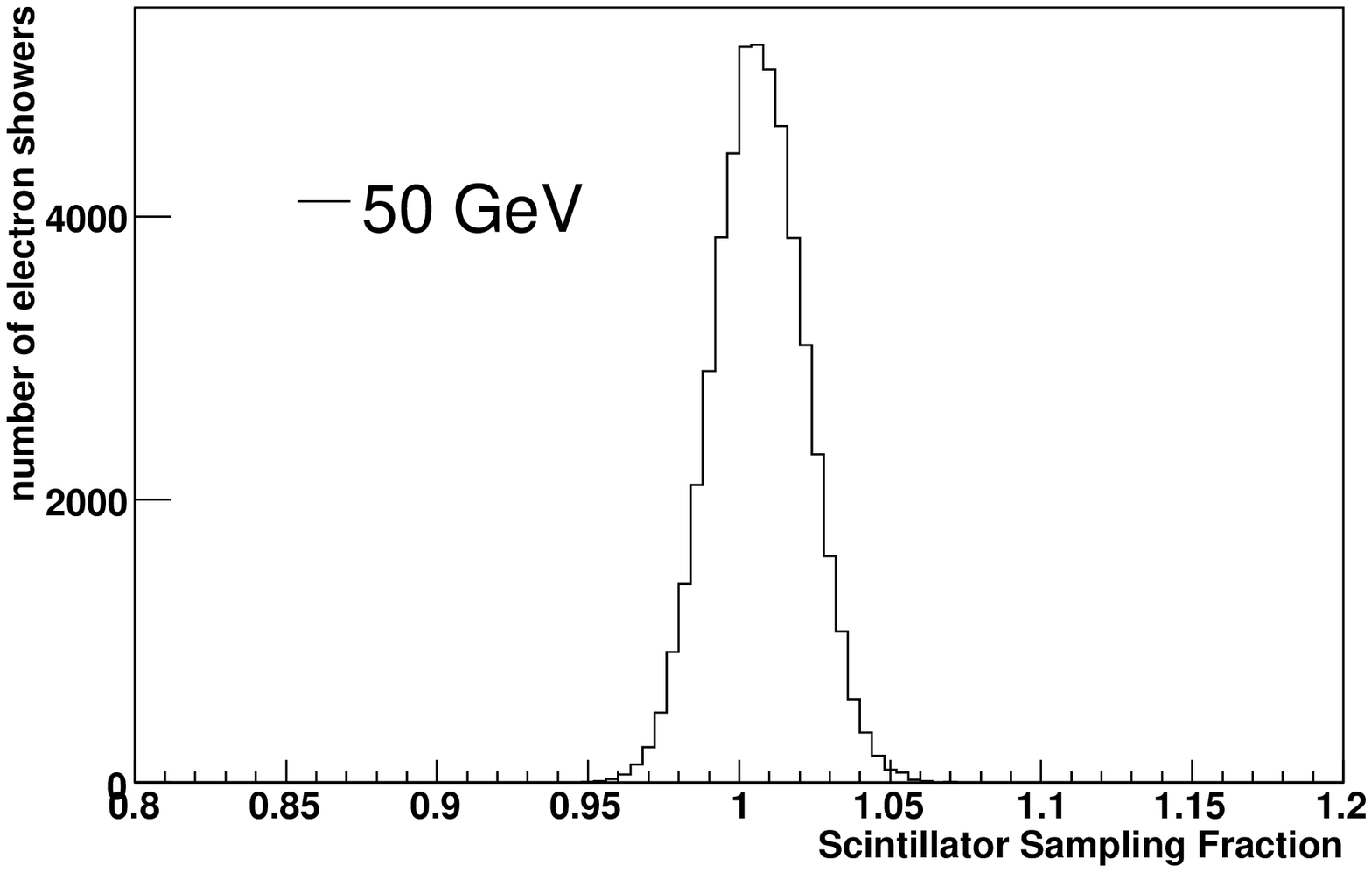}
\includegraphics[width=3.3in]{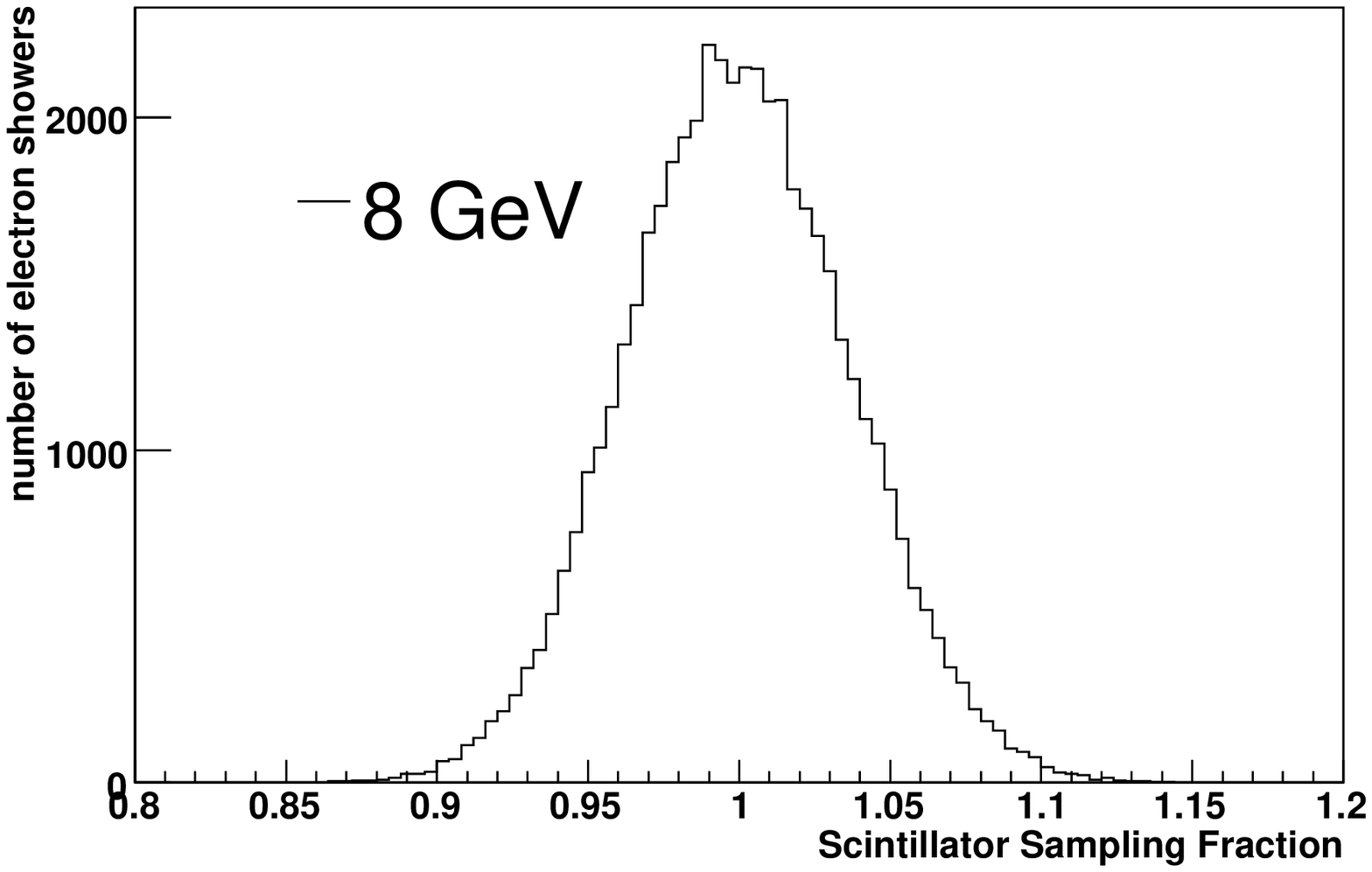}
\includegraphics[width=3.3in]{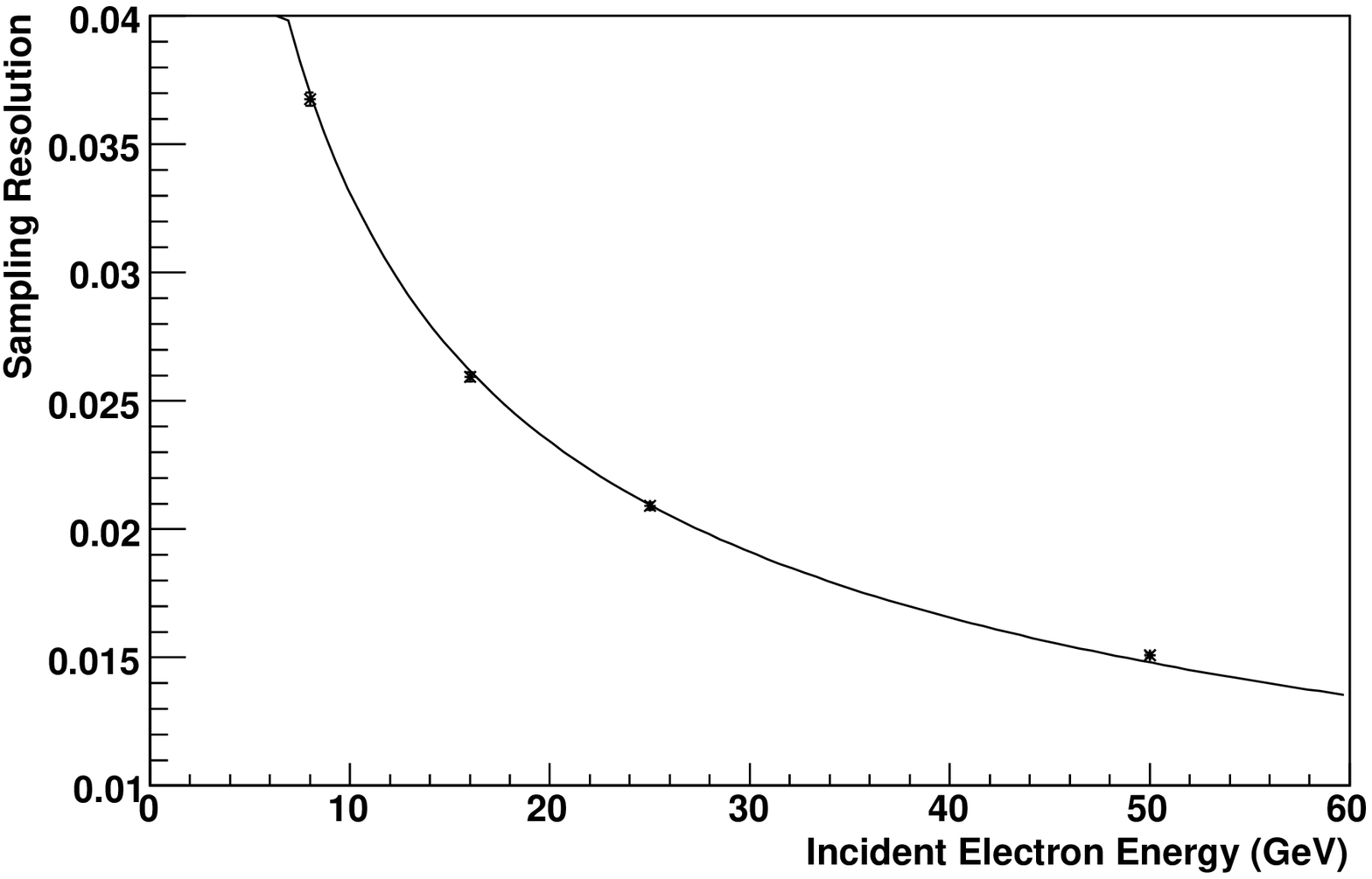}
\end{center}
\caption{Distributions of the scintillator sampling fraction $f_s$ for electrons with incident 
energy of 50 GeV and 8 GeV (top and middle).  The mean $f_s$ is normalized to unity. The lower plot
 shows the {\it rms} of the  $f_s$ distribution  for incident energies of 50 GeV, 25 GeV, 16 GeV and 8 GeV (points)
 with a fit of the form $k / \sqrt{E/{\rm GeV}}$ describing the sampling resolution.}
\label{samplingFractions}
\end{figure}

The fluctuations in $f_s$ correspond to the sampling resolution of the CEM, which is 
characterized as $\sigma(f_s) /f_s  = k / \sqrt{E_T/{\rm GeV}}$.  Figure~\ref{samplingFractions} 
shows the distributions of the normalized $f_s$ ({\it i.e.} $\langle f_s \rangle \approx 1$) at 
different energies.  To remove the effect of leakage on the resolution, these distributions 
are obtained from {\sc geant4} using a very deep calorimeter with no leakage.  To extract 
$k$ we plot the {\it rms} of each distribution as a function of incident energy.  A fit to the functional form  
$k / \sqrt{E_T/{\rm GeV}}$ yields $k=10.5$\%.  This calculation of the sampling resolution takes 
into account the fluctuations of the scintillator sampling fraction, but not the 
fluctuations of the scintillation  photons and photoelectrons in the phototubes.  These fluctuations  
contribute a sampling resolution of 7\%$/\sqrt{E_T/{\rm GeV}}$ based on the measurement of 200 
photoelectrons per GeV of incident energy~\cite{cemnim}.  Combining the two sources of 
sampling fluctuations in quadrature yields an effective sampling term of 12.6\%. 

\section{Sources of Non-Linearity}
\label{sec:nonlinearity}
A source of non-linearity at high energy is the longitudinal leakage from the back of the 
calorimeter, since the fraction of energy that leaks increases with energy.  The parameterization 
of longitudinal leakage was discussed in Section~\ref{sec:leakage}.  To isolate other sources 
of non-linearity, we eliminate the effect of longitudinal leakage by simulating a very deep 
calorimeter using {\sc geant4}. 
   
A deep calorimeter is expected to be 
 linear if on average a fixed (energy-independent) fraction of the total 
shower energy is deposited in the scintillator layers.  In the CDF detector, the uninstrumented 
material upstream of the CEM absorbs energy at the start of the shower without any measurement 
from a corresponding scintillator sampling layer.  This upstream 
 energy loss is an important source of 
non-linearity. 

This study encompasses several additional small sources of non-linearity.  First, a non-linear 
effect arises from the finite thickness of the lead absorber layers.  In the limit of very 
thin absorber layers, all shower particles generated in the absorber layers traverse the 
entire absorber layer and deposit some fraction of their energy in the subsequent scintillator 
layer.  In the CEM lead layers of 3.175~mm thickness, the typical ionization energy 
loss for an electron is 4.1~MeV per layer at normal incidence.  Soft photons that 
convert or Compton-scatter in the front part of a lead layer can generate lower-energy 
electrons which are fully absorbed in the lead, leaving no signal in the subsequent 
scintillator.  Thus, we expect a low-energy non-linearity in the calorimeter response due to 
the finite thickness of the lead absorber layers. 

\begin{figure}
\begin{center}
\includegraphics[width=3.5in]{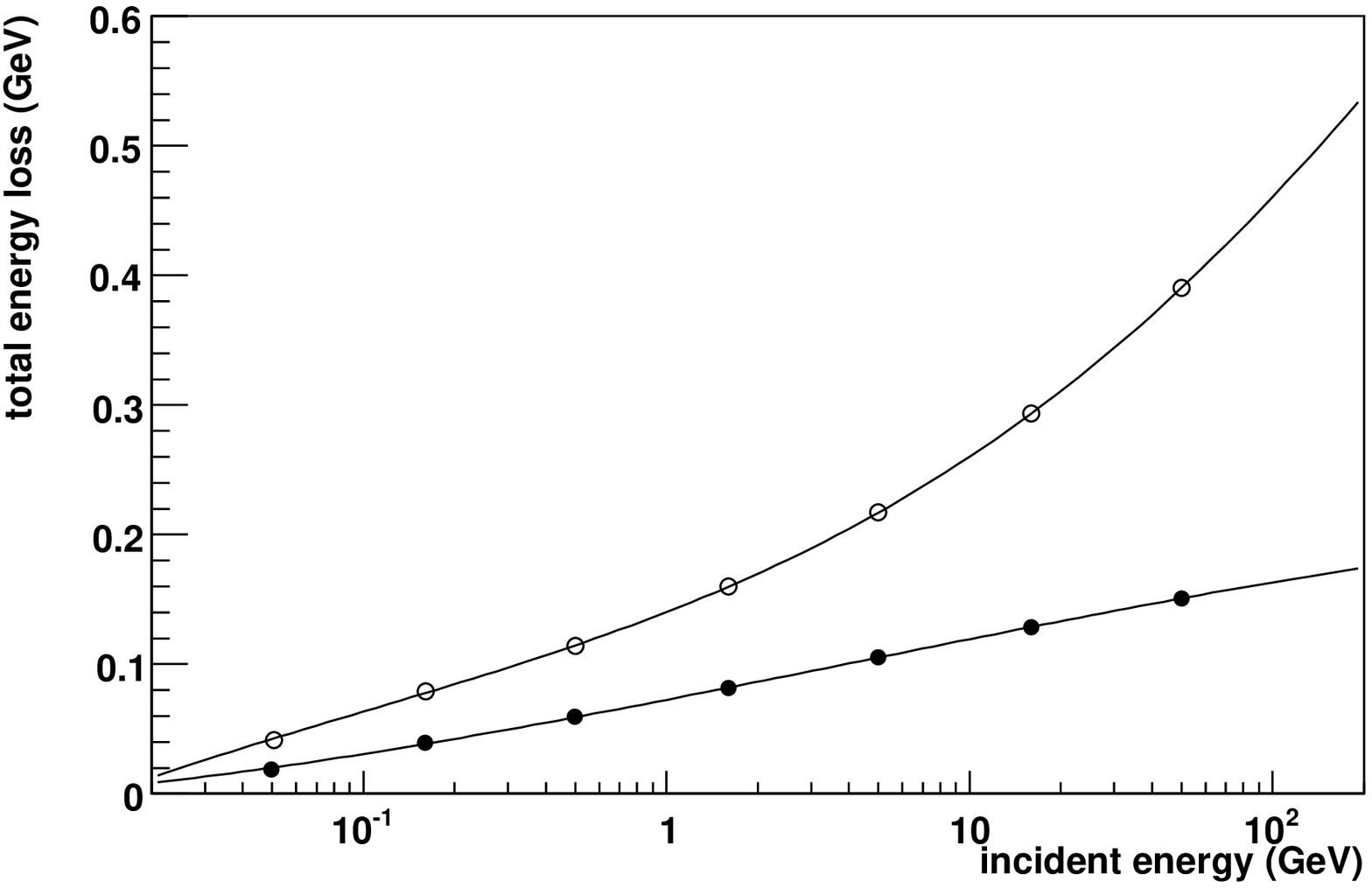} 
\includegraphics[width=3.5in]{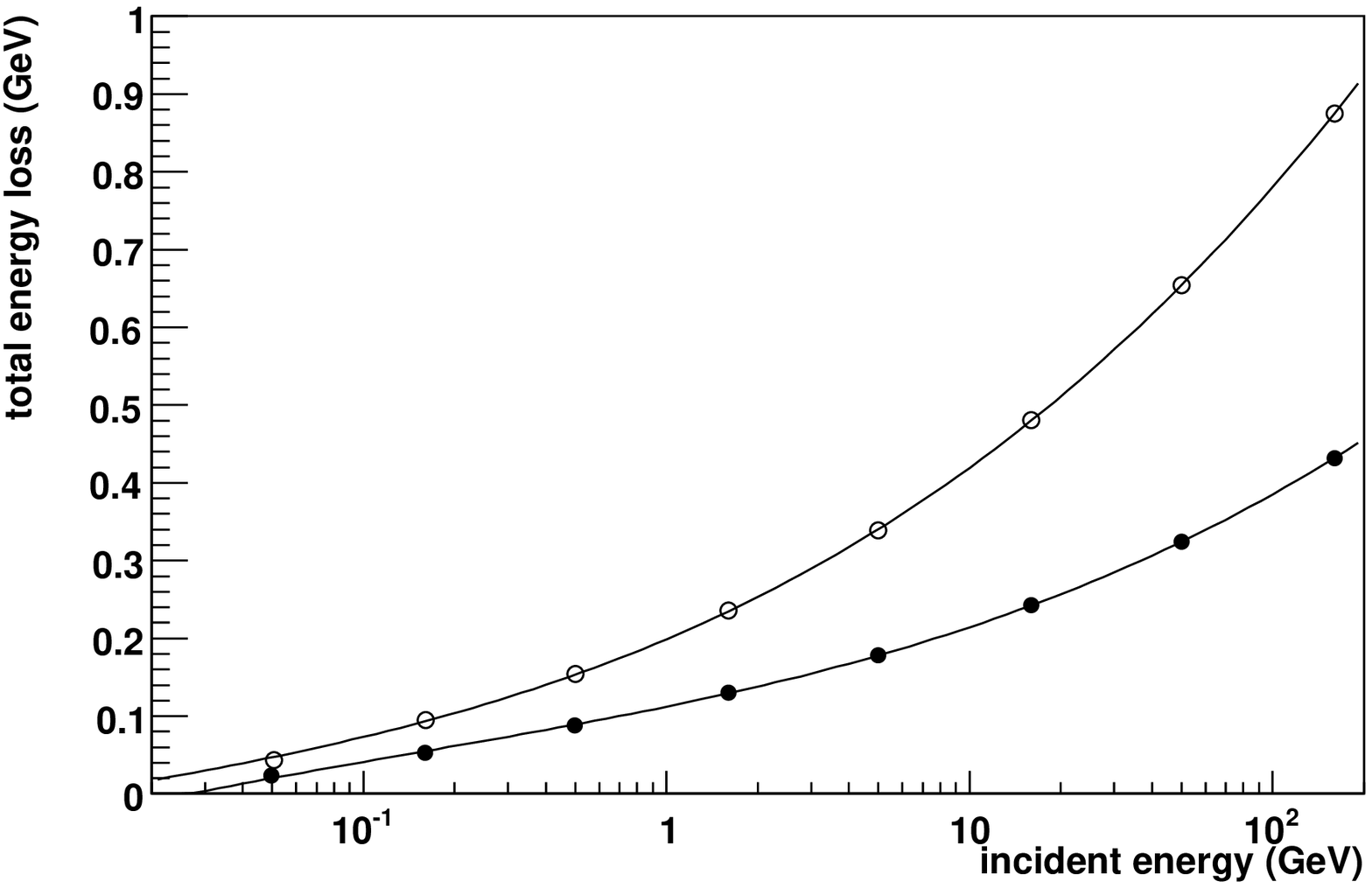}
\includegraphics[width=3.5in]{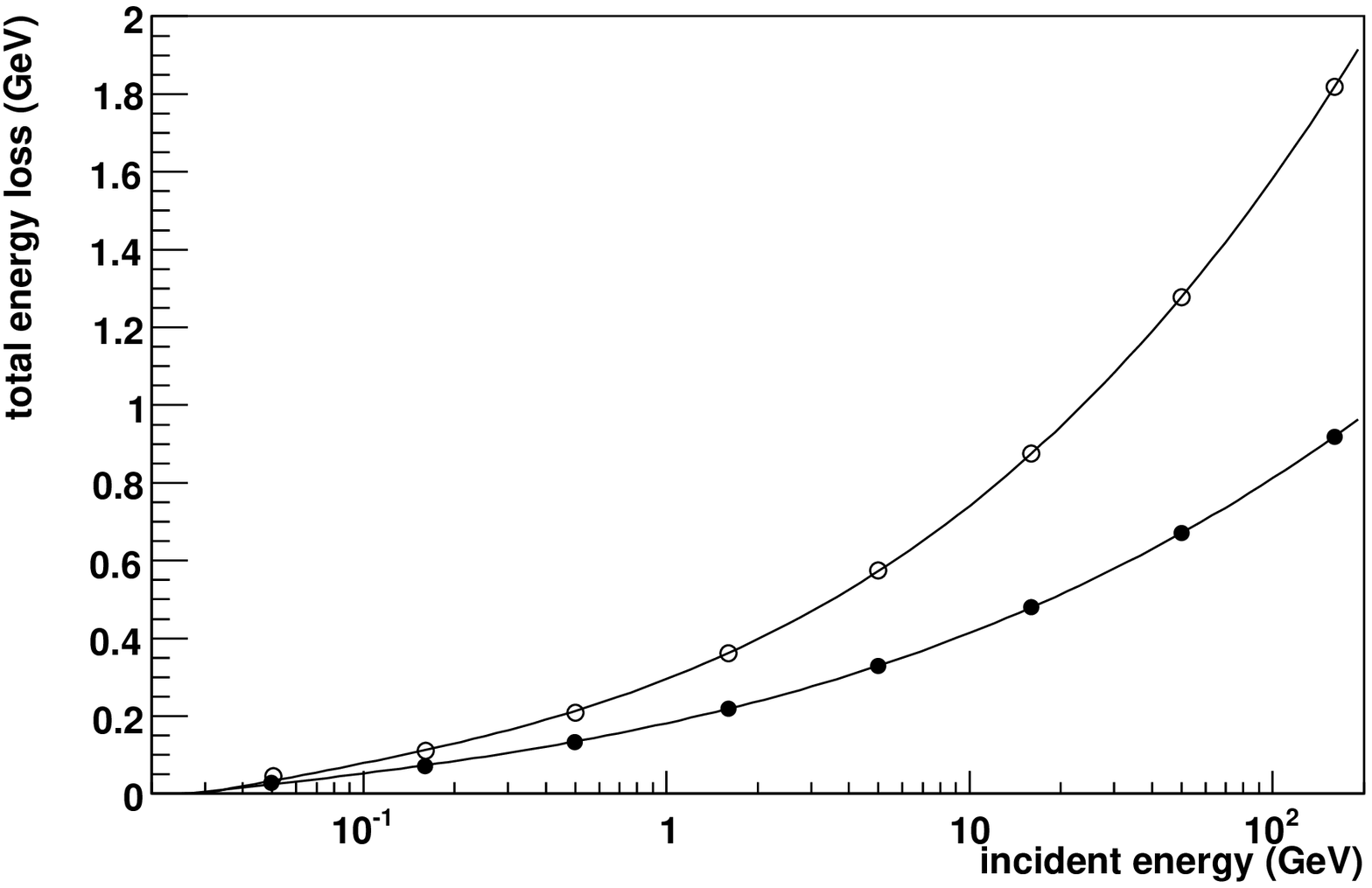}
\end{center}
\caption{The dependence of the energy loss $\delta$ on the energy of the incident photon (open circles) and 
electron (solid circles), at incident angles corresponding to $\csc \theta=1.16$ (top), 
$\csc \theta = 1.4$ (middle) and $\csc \theta =1.82$ (bottom).  The tower geometry 
corresponds to a very deep calorimeter of width 48 cm.  The smooth curve is a cubic 
polynomial fit in $\log_{10} (E_{\rm incident}/{\rm GeV})$. }
\label{deltaElectron4}
\end{figure}

Additional non-linearities could result from the transverse leakage of the electromagnetic 
shower or the back-scatter of soft photons.  The bulk of the unmeasured energy from these 
effects will be proportional to the incident energy.  However, a component of 
this energy loss may not scale with the incoming energy, thus causing a low-energy 
non-linearity.  

With incident electrons and photons, we use {\sc geant4} to calculate the energy deposited in 
the scintillators $E_{\rm s}$ as a function of the incident energy $E$. 
We define a low-energy non-linearity parameter $\delta$ as 
\begin{equation}
 \delta \equiv E - \frac{E_{\rm s}}{f} 
\label{deltaEqn}
\end{equation}
where $f$ represents a fixed scintilator sampling fraction.  The quantity $\delta$ has 
been defined such that ($E - \delta$) is strictly proportional to 
$E_{\rm s}$.  It represents an effective energy loss that can vary with energy, but 
we do not allow it to have a component proportional to energy, since this would simply 
redefine $f$. 

For this study the {\sc geant4} detector model contains a large number of layers to 
fully contain the shower longitudinally.  In the transverse directions we define the 
calorimeter to have dimensions of 48~cm $\times 48$~cm to emulate the size of a reconstructed 
electron cluster used in CDF data analysis~\cite{wmassPRD,xsecPRD}.  The {\sc geant4} prediction for $\delta$ is shown in 
Fig.~\ref{deltaElectron4} for incident photons and electrons.  Each point in the figure 
is evaluated by averaging over 50,000 showers at fixed incident particle energies ranging 
from 5 MeV to 160 GeV.  We see that $\delta$ is larger for photons than electrons of the same energy,
 and that $\delta$ increases logarithmically with incident energy.  This increase is well-described by a function of the form
\begin{eqnarray}
\delta = P_3(\log_{10}E) + \alpha E,
\end{eqnarray}
where $P_3(x)$ is a cubic polynomial in $x$.  The scintillator sampling fraction $f$ 
is iteratively tuned in Eqn.~\ref{deltaEqn} such that the fit returns $\alpha = 0$.

The non-linearity from the dominant sources -- uninstrumented material upstream of the 
CEM and the finite thickness of the absorber sheets -- tends to increase as the incidence 
angle of the electron or photon moves away from normal incidence.  This is because the 
increased path length through uninstrumented material causes more early showering and therefore 
more unsampled energy loss.  To demonstrate this dependence on incidence angle, 
Fig.~\ref{deltaElectron4} shows $\delta$ as a function of energy for $\csc \theta$ ranging 
from 1.16 to 1.82, where $\theta = 90^\circ$ corresponds to normal incidence.  The plots 
show that the energy loss increases with increasing $\csc \theta$. 

\section{Summary}
We have implemented in {\sc geant4} the longitudinal structure of the central electromagnetic 
calorimeter of the CDF detector, as well as the passive material between the outer edge of the 
tracker and the calorimeter.  Using this simulation, we calculate the longitudinal energy 
leakage for incident electrons as a function of incident energy and the total thickness of the 
structure in radiation lengths.  We parameterize this distribution and derive an additional 
parameterization for incident photons by converting each incident photon to an electron-positron 
pair.  We have also studied: the correlation between the energy deposited in the scintillator  and the longitudinal 
energy leakage; the scintillator sampling fluctuations; and the calorimeter non-linearity 
arising from upstream passive material and finite absorber thickness.  The parameterizations 
are used in the measurement of the $W$ boson mass with the CDF detector.

\vspace*{4mm}

{\bf Acknowledgements} \\
We wish to thank our colleagues on the CDF experiment for providing information on the construction of the detector. We thank Ravi Shekhar 
 for his assistance in installing the {\sc geant4} software.  
We acknowledge the support of the U.S. Department of Energy  
and the Science and Technology Facilities Council of the United Kingdom.


\begin{thebibliography}{99}
\bibitem{conventions}
Pseudorapidity is defined as $\eta = - \ln [\tan (\theta / 2)]$, 
where  $\theta$ is the polar angle from the beam axis. The azimuthal angle is denoted by $\phi$. Energy (momentum) transverse to the beam is denoted as $E_T$ ($p_T$). 
 We use the convention $c = 1$ throughout this paper.
\end{thebibliography}
\end{document}